\documentclass[referee]{aa}
\usepackage{epsfig}
\usepackage{graphics}
\usepackage{times}
\usepackage{supertabular}
\epsfclipon

\def\simgt{\rlap{\lower 3.5 pt \hbox{$\mathchar \sim$}} 
                                \raise 1pt \hbox {$>$}}
\def\simlt{\rlap{\lower 3.5 pt \hbox{$\mathchar \sim$}} 
                                \raise 1pt \hbox {$<$}}
\newcommand{\hea}{HE\,1122$-$1649$\;$ }
\newcommand{\heb}{HE\,0515$-$4414$\;$}
\newcommand{\ly}{\mbox{Ly$\alpha$}$\;$}
\newcommand{\dly}{damped Lyman $\alpha \;$}

\newcommand{\beq}{\begin{equation}}
\newcommand{\eeq}{\end{equation}}
\newcommand{\I}{\,{\sc i}$\,$}
\newcommand{\II}{\,{\sc ii}$\,$}
\newcommand{\III}{\,{\sc iii}$\,$}
\newcommand{\IV}{\,{\sc iv}$\,$}
\newcommand{\V}{\,{\sc v}$\,$}
\newcommand{\VI}{\,{\sc vi}$\,$}
\newcommand{\nh}{\mbox{$N$(H\I)}$\,$}

\newcommand{\km}{km\,s$^{-1}$\,}

\begin{document}
\thesaurus{03(11.17.1; 11.17.4 HE 1122-1649, HE 0515-4414; 11.01.1;
  11.05.2; 11.09.4; 12.03.3)}
\title{Damped Lyman \boldmath $\alpha$ \unboldmath systems at 
\mbox{ \boldmath $z=0.68 $ \unboldmath} and 
\mbox{\boldmath $z=1.15 $ \unboldmath} towards 
\hea and \heb 
\thanks{Based on observations performed at ESO/La Silla,
on observations with the NASA/ESA Hubble Space Telescope 
and on observations at Keck Observatory.}}
\author {A.~de la Varga \inst{1}
\and D.~Reimers \inst{1}
\and D.~Tytler \inst{2}
\and T.~Barlow \inst{3}
\and S.~Burles \inst{4}}
\institute{Hamburger Sternwarte, Universit\"at Hamburg, 
Gojenbergsweg 112, D-21029 Hamburg 
\and Center for Astrophysics and Space Sciences,
 University of California, San Diego, MS-0424, La Jolla, CA 92093-0424
\and California Institute of Technology, 100-22, Pasadena,
 CA 91125
\and Department of Astronomy and Astrophysics, University of Chicago,
5640 S. Ellis Ave., Chicago, IL  60637
} 

\offprints{A.~de la Varga}
\mail{avarga@hs.uni-hamburg.de}

\date{Received date/ Accepted date}

 \maketitle

\begin{abstract}

Detailed analysis of the spectral lines from two low redshift damped
Lyman alpha absorbers (DLA) confirms that they are a heterogeneous
population.  Both systems have low metal abundances of approximately
0.05 to 0.1 solar.  The abundance pattern of the DLA at $z=1.15$
towards HE\,0515-4414 shows dust depletion comparable to that found in
our Galaxy, while the system at $z=0.68$ towards HE\,1122-1649 shows
abundance ratios resembling metal-poor halo stars with no dust
depletion, for a comparable H\I column.  Constraints for N/Fe and N/Si
also hint at $z=0.68$ DLA as a galaxy with recent star formation.  
The trend of increasing C\I/H\I with decreasing $z$ is discussed. 
Only weak absorption from highly-ionized species associated with the
DLA at moderate $z$ have been detected and it probably originates in regions
distinct from the low ionization gas.  The low-ion profiles show very
complex structures and are too large to be explained by the rotation
of a disk: the system at $z=0.68$ spans over $\sim$ 300 \km and the
$z=1.15$ DLA presents substructure over more than 700 \km, the largest
velocity extent found up to date for a DLA.


\end{abstract}

\section{Introduction}

Damped Lyman $\alpha$ absorbers (DLA) are the population of 
quasar absorption systems showing the highest neutral
hydrogen column density (\nh $\ge 2\times 10^{20}$ cm$^{-2}$).
At high redshift
they contain the bulk of the cosmological mass of neutral gas in the
Universe and are believed to be progenitors of current
galaxies (e.g. Wolfe et al. 1986), although their nature and properties 
still remain unclear.
They provide information about the chemical enrichment history of
galaxies which can be contrasted with the studies of stellar
populations in the local galaxies.
Several samples of DLA at $z > 1.7$ have been 
studied thoroughly
(e.g Pettini et al. 1990, Lu et al. 1996, Pettini et al. 1997a,b)
but very few systems are known at intermediate or low redshift (Pettini \&
Bowen 1997, and references therein; Boiss\'{e}. et al. 1998, 
de la Varga \& Reimers 1998, 
Pettini et al. 1999a,b). 
To interpret the high-$z$ data correctly, 
it is essential to follow the evolution of the DLA population 
to the present time.
Moreover, at low $z$ it is possible to search for luminous counterparts
via imaging (e.g. Le Brun et al. 1997, 
Steidel et al. 1994, 1995, 1997, Rao \& Turnshek 1998).
 
We herein present observations of two DLA, at $z=0.68$ and $z=1.15$,
in the line of sight to \hea and \heb, respectively.
The combination of HST spectra with high resolution spectra
(CASPEC and HIRES) allows a detailed study of metal abundances and
kinematics at low redshift.

Total metal abundances provide information about the 
population of absorbers which constitute DLA.
Measurements of Zn and Cr show
abundances of 1/15 solar, but with a considerable spread 
(Pettini et al.\ 1997b). 
The metallicity is not significantly higher for the few systems found
at $z<1.5$ (Pettini et al. 1999a,b).  
The underabundance of Cr relative to Zn 
(Pettini et al.\ 1997a, 1999a,b) suggests that dust is also present, 
though the depletion level seems to be lower than in the
Galaxy. The same conclusion was drawn from reddening studies of
QSOs with and without DLA in their line of sight 
(e.g. Pei et al. 1991).
The interplay between nucleosynthesis and dust depletion from the
study of relative abundance patterns in DLA is still unclear
(cf. Lu et al. 1996, Prochaska \& Wolfe 1996, Pettini et al. 1997b,
Kulkarny et al. 1997)
and, however, crucial to understand the evolution and chemical enrichment of
galaxies.  

Studying the kinematics of the absorption line profiles should
clarify the nature of DLA and could help to discern among theories of
galaxy formation: are DLA related to rotating disks or rather to irregular
protogalactic clumps merging, as expected in scenarios of hierarchical
cosmogony? (cf. Prochaska \& Wolfe 1997a,b,1998, Haehnelt et al. 1998,
McDonald \& Miralda-Escud\'{e} submitted).
Comparing profiles at low and high $z$ could help to
understand whether the asymmetries arise from a mixture of kinematical
structures (random motions, rotation, infall, and merging...), or
whether a trend can be recognized.

Our paper is organized as follows: in Sect.~2 we describe the observations
and the data reduction. The analysis of the absorption lines
is discussed in Sect.~3,  the metal abundances 
in Sect.~4, and  the relative abundances in Sect.~5.
The neutral carbon and molecular content are studied in Sects.~6 and 7, 
respectively.
In Sect.~8 the kinematics of the absorption profiles is 
considered, and  in Sect.~9 we study the highly-ionized gas
related to the systems. 
Finally, concluding remarks are summarized in Sect.~10.

\section{Observations and Data Reduction}

\subsection{\boldmath {\em HST} \unboldmath Spectra}

Ultraviolet spectra of \hea ($z_{em}=2.4$, V=16.5, Reimers et al. 1995) 
were obtained with the Faint Object
Spectrograph (FOS) onboard the Hubble Space Telescope (HST) in 1995. 
With gratings G270H and G190H (red
digicon and entrance aperture of $3.7^{\prime\prime}\times
3.7^{\prime\prime}$),
the wavelength range from 1600 to 3280 \AA\ was covered 
with a maximum signal-to-noise ratio of 28.
The spectral resolution is FWHM = 1.4 and 2.0 \AA\ for 
G190H and G270H, respectively (Schneider et al. 1993).
The wavelength calibration of the pipeline reduced data was examined
by comparing the overlap region between the G190H and the G270H spectra,
and also with the Keck spectra, 
which have an overlap of 100 \AA\ with the FOS spectra.

The spectra were corrected for interstellar reddening using the
Seaton law (Seaton 1979) with $E(B-V)=0.08$ , corresponding to
\nh=4.16$\times10^{20}$ cm$^{-2}$.  
The high density of absorption lines complicates the measurement of a
local continuum. 
Regions apparently free of absorption were selected 
by checking the error in the mean flux for consistency with the noise,
and cubic splines were fitted to the mean flux values in
those regions.

\subsection{Keck \boldmath {\em  HIRES} \unboldmath Spectra}

Optical data of \hea were taken with the HIRES
spectrograph (Vogt et al.~\cite{vog94}) on the W.M. Keck1 10-m telescope
with a total of 3 hours exposure in April 1997.
The spectral range was from 3150 to 4710 \AA\ with a S/N
of 19/35 at 3500/4500 \AA. 
We used a 1.14" x 7.0" slit,
which gave a resolution of 8 \km FWHM (R=37500) and adequate sky coverage,
and used the UV cross-disperser
to increase efficiency at wavelengths below 4200 \AA.
The images were processed and the spectra were
optimally extracted using an automated
package specifically designed by T. Barlow for HIRES spectra.
Thorium-Argon lamp images were obtained immediately after the observations
to provide wavelength calibrations in each echelle order.
The root-mean-square residuals in the wavelength calibration for each
echelle order were less than 0.3 \km.
All wavelengths are vacuum values in the heliocentric frame.
Each echelle order was continuum fit with a Legendre polynomial
to normalize the intrinsic QSO flux to unity.

\subsection{\boldmath {\em CASPEC} \unboldmath Spectra}

Further optical spectra were obtained with CASPEC (Long Camera) 
at the 3.6\,m telescope at ESO/La Silla. 
The data reduction was done with the ECHELLE software
program available in MIDAS, supplemented by software programmes
for an optimal extraction of the echelle orders developed by
S.~Lopez.
For wavelength calibration purposes,
Th-Ar spectra were taken immediately before and after each quasar
observation. The uncertainty in the calibration, $\Delta \lambda$, is
given by the rms residuals after fitting third order polynomials 
to the lines identified 
in the arc lamp observations. For all  the CASPEC observations  
we obtain $\Delta \lambda \simlt 0.2$ of a resolution element.
The wavelength scales were corrected to vacuum-heliocentric values 
and individual observations rebinned to the same wavelength scale
and combined by weighting by the inverse variance.
The continuum for each echelle order was determined by iterative cubic 
spline fitting to regions free of absorption lines.

{\em \hea:} \\
\noindent
In the observations of \hea in 1994, 43 echelle orders covered the wavelength
range from 3750 to 5250 \AA. The achieved
resolving power, measured from the emission 
lines of the Th-Ar exposures, 
was R=17\,000, and the signal-to-noise ratio for the combined spectrum 
was 34, 43 and 53 
at $\lambda_{\rm obs}$= 4000, 4500 and 5000 \AA, respectively.
In 1996, 48 orders covered from 3320 to 4600 \AA\ with S/N of 8/29 at 3500/4500
\AA\ and R=20\,000.

{\em \heb:} \\
\noindent
We observed \heb using CASPEC in October 1996 
  from 3530 to 4830 \AA\ with R=21\,500,
and 4865 to 6200 \AA\ with R=24\,000. Further observations 
in November 1997
provided us with the ranges from 3250 to 4560 \AA\ (R=27\,000), 
5430 to 7870 \AA\ 
(R=28\,600) and 6140 to 8570 \AA\ (R=29\,000). The combined spectrum
covering the spectral range from 3250 to 8570 (with a small gap
between 4830 and 4865 \AA)
has a maximum signal-to-noise ratio of $\simeq 55$.

\section{Analysis of the Absorption Lines}

Using the FITLYMAN program in the MIDAS package (Fontana \& Ballester
1995), the absorption lines were fitted with Voigt profiles convolved
with the instrumental profile, simultaneously determining the
redshift, column density, $b$ values and associated errors, while
minimizing a $\chi^2$ parameter matrix.
Line parameters were taken from Morton (1991) and from Verner et al. (1994).
Different transitions from the same ion were fitted simultaneously, and 
different ions and elements were fitted independently.
We chose the minimum number of components which gave an adequate fit.
Several fits were performed for the same
complex by shifting slightly the initial $z$ values and testing new sets. 
The considerable arbitrariness in the choice of components 
does not seriously affect the total column densities derived
for most ions, as the presence of several transitions 
allow us to discern blends and hidden components --at least  
in the high resolution spectra-- by means of the apparent optical depth
method (Savage \& Sembach 1991).  

\subsection{DLA at $z=0.68$  towards HE\,1122$-$1649}
\begin{figure}[ht]\centering   
\epsfig{figure=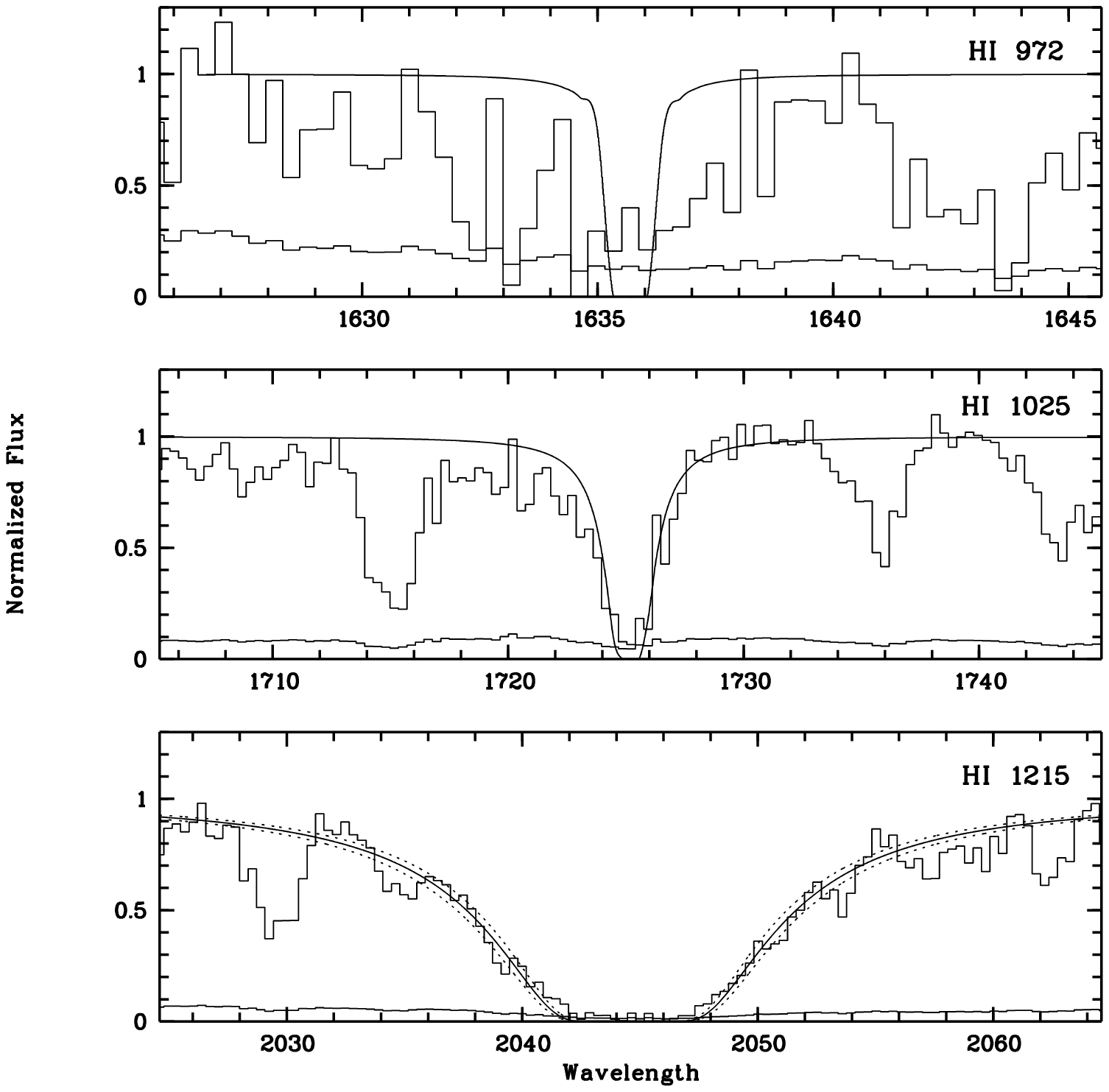,width=11cm}
\caption{\label{he1122_hI}{
HI Ly$\alpha$, Ly$\beta$ and Ly$\gamma$
for the damped Lyman $\alpha$ system at $z=0.6819$ towards \hea.
The best fit is found for log\nh$=20.45\pm 0.15$ fitting   Ly$\alpha$ 
and Ly$\beta$ simultaneously (Ly$\gamma$ was not used, since it lies
on a too noisy and blended region).}}
\end{figure}

The profile of the \dly is well fitted in the FOS spectrum by a column
density of neutral hydrogen of log\nh $=20.45 \pm 0.15$ at $z=0.6819$
 (Fig.~\ref{he1122_hI}).  A small correction for scattered light
($\approx 1.5$ \% of the continuum) was applied to bring the core of
the absorption line to zero flux.

\begin{figure}
\vspace{0cm}
\hbox{\hspace{0cm}\epsfig{figure=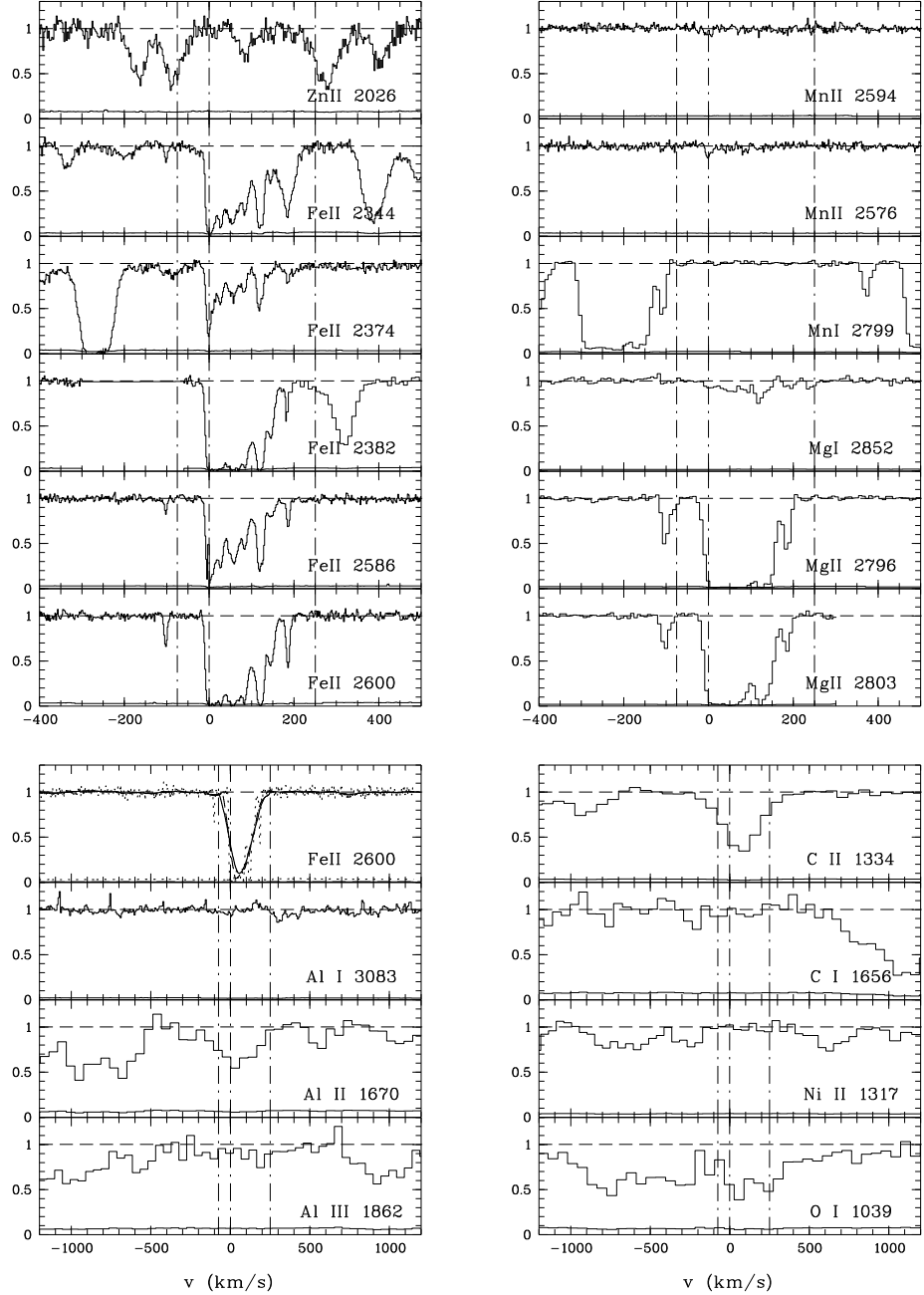,width=15cm}}
      \caption{\label{ions_1122A}{ Velocity profiles of some ions associated with the DLA
at $z=0.68$ towards \hea ($v=0$ \km corresponds to $z=0.68189$).  The
solid line overplotted for the ion Fe\II 2600 in the bottom-left panel
shows the absorption profile as it would be seen at the HST/FOS
resolution (compare with Fe\II 2600 at Keck/HIRES resolution in
upper-left panel)}}
   \end{figure}

\begin{figure}
\vspace{0cm}
\hbox{\hspace{0cm}\epsfig{figure=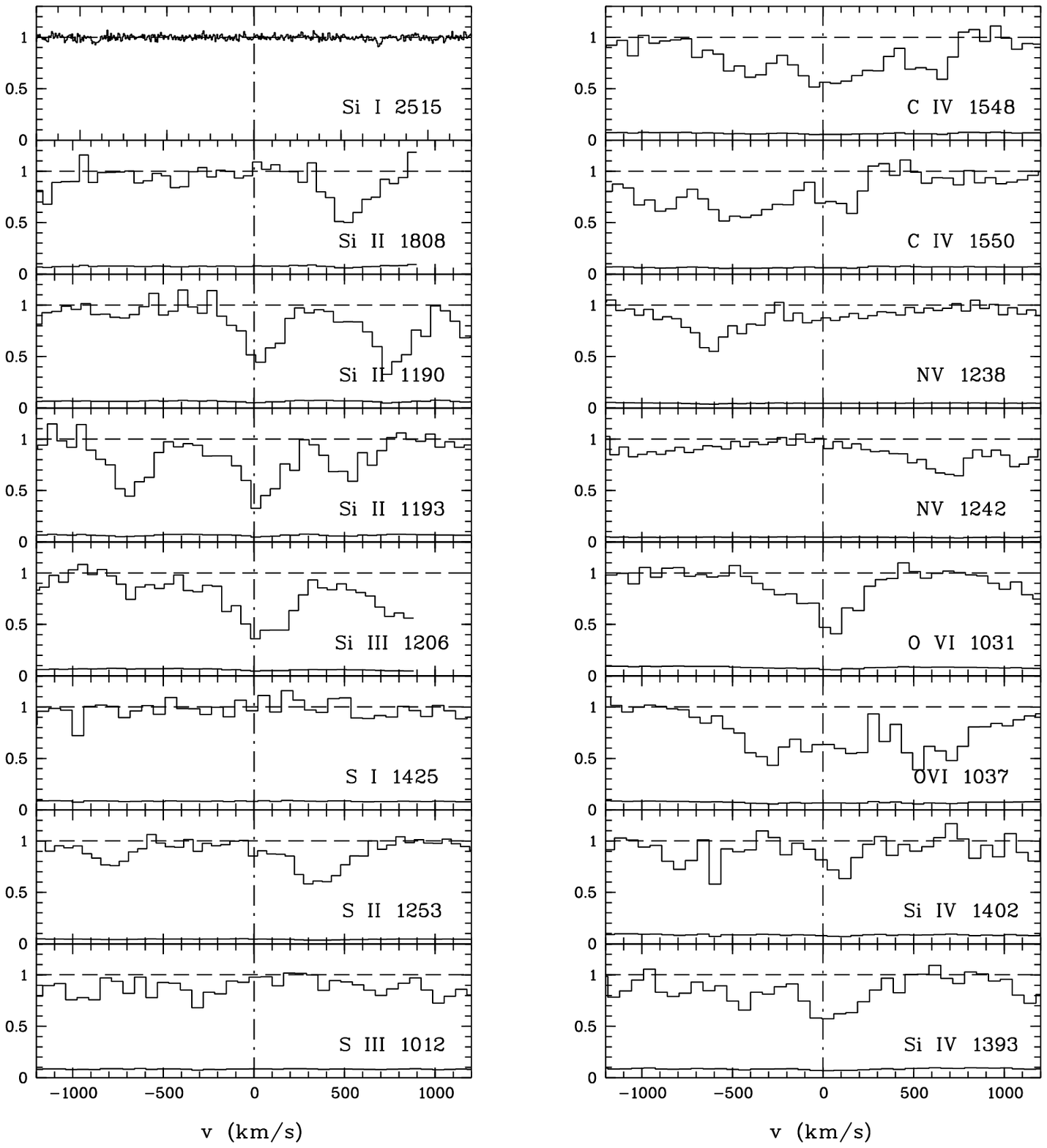,width=15cm}}
\vspace{0cm}
      \caption{\label{ions_1122B}{ 
(...) Velocity profiles of some ions associated with the DLA at $z=0.68$
towards \hea ($v=0$ \km
corresponds to
$z=0.68189$).
}}
   \end{figure}

We detect absorption at the expected position of numerous 
metal lines associated with this system.
Table~1 lists the line parameters resulting from the fits,
and Fig.~\ref{ions_1122A} and \ref{ions_1122B}   show
some of the absorption features in velocity space. 

We adopt $3\sigma$ upper limits for lines which were not detected.

Comparable $b$ values resulting from fits of individual components for
different ions suggest that turbulent broadening is dominant.
Under this assumption, for the cases where blending 
complicated the fitting, a typical $b$ value was adopted 
as derived from the high resolution spectra for resolved lines. 
Keck spectra were filtered at the FOS instrumental
resolution and new fits of the Fe\II transitions longwards of \ly
were performed.
We obtain $b$ values of the order of $\approx 30$ \km by
fitting the whole complex as a unique absorption feature at the
resolution of the G270H spectra, and $b \approx 25$ and 16 \km fitting with two
components at the resolution of G190H. 
Because of the severe blending in the \ly forest, the line fits performed
at the limited FOS resolution should be considered  upper limits.
However, the study on metal abundances will be 
based on column densities derived from the high resolution spectra. 

The  numerous Fe\II transitions 
observed in the Keck spectra allow a reliable estimation of N(Fe) by fitting the
strongest components in non-saturated transitions
(Fe\II2260,2374). The best fit for Fe\II\ requires at least 14
components (13 in the main feature plus the satellite component at 
$ v \simeq -100 $ \km).

The non-detection of Zn\II 2026
indicates that logN(Zn) $< 11.76$. 
Several tests that modified the continuum level locally
by means of fitting adjacent \ly lines did not alter significantly 
the quoted value for the upper limit.
Cr\II transitions are blended with
\ly lines and with C\IV at $z=1.234$, so that an estimate of the column
density for Cr\II could not be derived. 
For the absorption feature between $ v \simeq 100$ and 170 \km, 
Cr\,{\sc ii}\,2062 yields an upper limit of logN(CrII)$\leq$ 12.2
while logN(ZnII)$\leq$ 11.8.
Ti\II 1910.6,1910.9 are in a blended and noisy region of the
spectrum, but recent Keck spectra (kindly provided 
by A. Songaila) covering up to 6060 \AA\  allowed the
detection of Ti\II 3384 (at $4\sigma$ level) at $z=0.6818927$ with
logN(Ti\II)$=11.79 \pm 0.28$, $b=1.22 \pm 0.77$.

\subsection{DLA at $z=1.15$  towards \heb}

\begin{figure}
\vspace{0cm}
\hbox{\hspace{0cm}\epsfig{figure=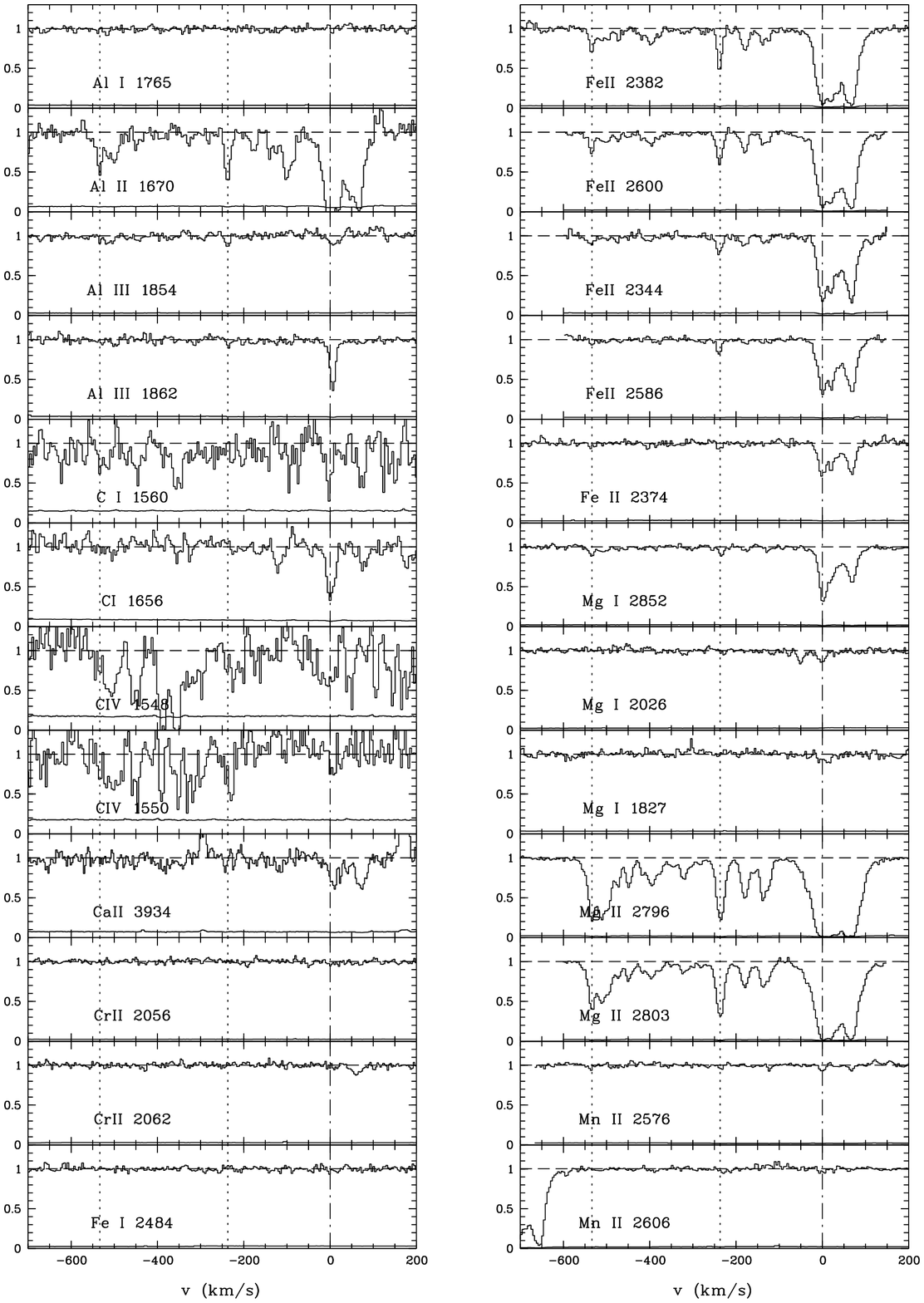,width=15cm}}
\vspace{0cm}
      \caption{\label{ions_0515A}{Velocity profiles of some ions
associated with the DLA at $z=1.15$ towards \heb ($v=0$ \km
corresponds to $z=1.15080$).}}  
\end{figure}

\begin{figure}
\vspace{5.cm}
\begin{center}
\epsfig{figure=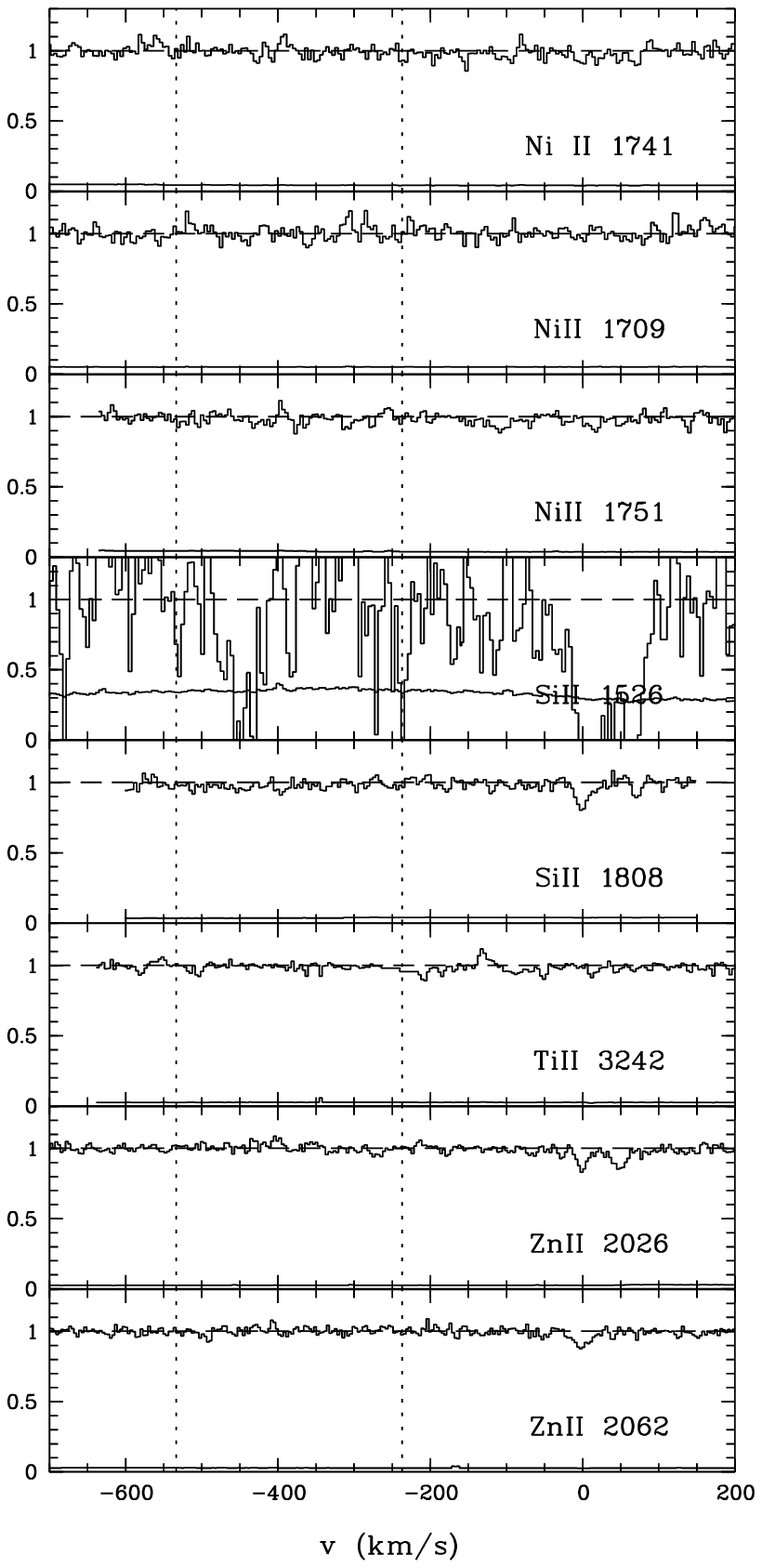,width=7cm}
\vspace{0cm}
\caption{\label{ions_0515B}
{(...)Velocity profiles of some ions associated with the
DLA at $z=1.15$ towards \heb ($v=0$ \km corresponds to $z=1.15080$).}}
\end{center}

   \end{figure}

According to IUE data, the extremely bright QSO HE\,0515$-$4414
($z$=1.7, V=14.9, Reimers et al. 1998) has a DLA at $z$=1.15 with an
estimated column density for neutral hydrogen of 
logN(H\I) $\simeq 20.4$.
We observe absorption associated with this system  at the
expected positions of Al\II 1670, Al\III 1854,1862, C\I 1656, C\IV
1548,1550, Ca\II 3934, Fe\II 1608,2344,2374,2382,2586,2600, Mg\I
1827,2026,2852, Mg\II 2796,2803, Mn\II 2576,2594,2606,  Ni\II 1741,
Si\II 1526,1808, Zn\II 2026,2062. Upper limits for the column density
of ions not detected over $3\sigma$ are determined for Cr\II
2056,2062, Fe\I 2484, Na\I 3303, S\I 1807 and Ti\II 3242.
Figs.~\ref{ions_0515A} and \ref{ions_0515B} 
show some of these ions in  velocity space, 
and Table~3 the line parameters
resulting from the fits.  The line profiles of transitions of Mg\II,
Fe\II and Al\II reveal that this remarkably complex system extends over
$\sim 700$ \km. 
At least three substructures can be distinguished: from
approximately $-575$ to $-270$ \km, from $-260$ to $-100$ \km and the main
absorption feature from $-80$ to 140 \km. 
Table~4 shows the derived column densities and 
metal abundances.
The number of components is possibly underestimated because of the
blending and blanketing problems.  The $b$ values resulting from the
fit of a blend cannot be well determined, but the column density of a
blend should be a good approximation of the column densities of
individual lines, when the system is not heavily saturated (Jenkins
1986).  The best fit for 
Mg\II and Fe\II requires at least 20
components for the whole system.

The main features in the Fe\II\ complex were fitted on the weakest
transitions (unsaturated), and the secondary ones on the strongest
transitions (Fe\II 2382,2600 were used for the absorption from $-575$
to $-270$ \km and Fe\II 2382,2600,2344 for the range $-260$ to $-100$
\km).  
Al\II 1670 is saturated in the strongest
substructure, and blended with Mg\II at $z=0.2813$ in the secondary
one. Al\III 1862 is blended with Mg\II 2803 
at $z=0.4291$.

A marginal $2\sigma$ Ni\II detection was obtained 
summing the spectral regions of 
Ni\II 1741, Ni\II 1709 and Ni\II
1751 to improve the signal to noise ratio
(stacking technique, e.g. Lu 1991).
The upper limit for N(Cr) was derived in the same way.

\subsection{Mg\II System at $z=0.28$ towards \heb}

\begin{figure}
\begin{center}
\epsfig{figure=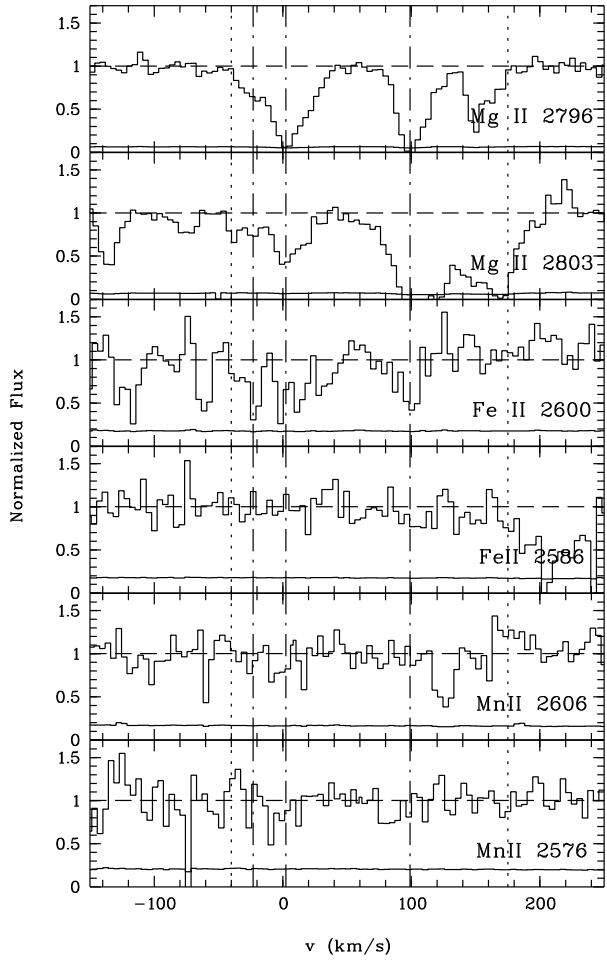,width=12cm}
\vspace{0cm}
\caption{\label{candidate}{Velocity profiles of ions 
associated with the Mg\II system at $z=0.28135$ 
(corresponds to $v=0$ \km ) towards \heb.}}
\end{center}
   \end{figure}

The strong Mg\II 2796 absorption features at $z=0.281$ reveal the
presence of a multicomponent system spanning over $\sim $200 \km 
from $-40$ to 175 \km, if $v=0$ \km corresponds to $z=0.28135$ 
(Fig.~\ref{candidate}).
The corresponding Mg\II 2803 lines could only be seen for
the main structure (from $-50$ to 40 \km), since the rest is blended 
with Al\II 1670 at $z=1.15$. The low redshift does not allow
to identify many associated features in our optical data.
Absorption lines at the expected positions of Fe\II 2600, 2586,
Fe\I 3021, Mn\II 2576,2606, Ca\II 3969 and Al\I 3803 have been looked
for and the results of the fits 
are shown in Table~6. 
We obtain a Mn\II/Fe\II ratio about 15 times larger than solar,
an improbable high value for a possible DLA. 
This suggests that the ionization is more complex, 
and/or that the continuum in the noisy region where Fe\II is expected 
could be underestimated. 
Several tests considering higher local continua yield an upper
limit of N(Fe)$\leq 13.7$ which would mean an abundance of Mn 
relative to Fe  five times
larger than solar, if the Mn\II 2576 feature is real.
It remains unclear whether this system is a new DLA
candidate at very low redshift, maybe with an important part of Fe\II
depleted onto dust.
Based on  N(Mn\II) and assuming a Mn abundance relative to H of $1/10$
solar, 
we obtain 
log\nh $\simeq 19.9$.

\section{Total Metal Abundances}

In Table~2, 4 and 5 we show the total metal abundances derived
assuming that H\I dominates and,
therefore, that ionization corrections are not important,
as shown by ionization models (Lu et al. 1995, Viegas 1995).

Both DLA are metal-poor with values for the metallicity
of 1/10 to 1/20 solar.
These values are comparable to the abundances found in 
other DLA 
and confirm the results from previous studies
(Lu et al. 1996, Pettini et al. 1997a,b, 1999a,b
and references therein.):\\
 {\em  (i)} The large scatter in the
 metallicity of DLA at any redshift 
 (1.6 in [Zn/H] over the whole range of $z$) 
is consistent with the heterogeneous types of galaxies seen in 
the cases  where the
associated galaxy could be identified (e.g. Le Brun et al. 1997,
Steidel et al. 1994, 1995, Rao \& Turnshek 1998).
Different star formation rates in spiral galaxies  
(Lindner et al. 1999),  different impact parameters, and possible local
inhomogeneities, would contribute to the scatter.\\
{\em  (ii)}
The enlarged sample shows no metallicity
increase with decreasing $z$, contrary to expectations from models of
cosmic chemical enrichment (e.g. Madau et al. 1996).
Including the new values in the sample
compiled from the literature, 
we obtain a column density-weighted mean
abundance of Zn (Pettini et al. 1997a) of
$\langle$[Zn/H]$\rangle \leq -1.05$ at $\langle z \rangle = 1$,
or $\langle$[Zn/H]$\rangle _{\rm corr}\leq -0.9$ if Zn is weakly depleted
onto dust as in the Galaxy.

A bias in the observations against dust-rich
absorbers could mean that 
the DLA found at low and moderate $z$, 
 where the dust presence should be more important, are not
 representative galaxies (e.g. Boiss\'{e} et al. 1998 and
ref. therein).
However, even considerable amounts of dust would not have been able to
obscure the line of sight of the bright \heb, and the intervening DLA
still shows a rather low metallicity [Zn/H]$\simeq -1$.  The upper
limit of [Zn/H]$<-1.3$ at $z$=0.68, obtained at the Keck resolution,
is also noteworthy.

\section{Relative Metal Abundances}

The ratio of two elements measured in DLA can vary from the solar
value due to a different intrinsic pattern and/or to the condensation
of refractory elements into dust depletion.  We expect that the DLA
might have abundance patterns resembling those in Halo stars, which
show an overabundance of $\alpha$-group elements relative to Fe, and
the odd-even effect (e.g. Mn underabundant relative to Fe.)

Assuming that elements in DLA are depleted as in the Galactic
interstellar medium, Vladilo (1998) corrected the observed relative
abundances from dust effects. He found the 17 systems of his sample
chemically homogeneous over a range of metallicities from nearly solar
down to $1/32$ solar and the dust-corrected abundances remarkably
close to the solar values. 
Applying the same correction to our systems, 
we obtain the dust-corrected abundances shown in Table~\ref{ratios}.

Due to the probable heterogeneity of the DLA population, 
it is interesting to discuss the abundance pattern in the systems
individually.

\subsection{DLA at $z=1.15$ towards HE\,0515$-$4414: evidence for dust depletion}

The refractory elements Cr, Ti, Ni are strongly depleted in this
system: [Cr/Zn]$<-1.04$, [Ni/Zn]$<-1.19$, [Fe,Ti/Zn]$<-0.66$.  These
abundances are comparable with those observed in the warm disk and
halo of our Galaxy (Savage \& Sembach 1996) .
The {\em corrected} ratios are, in general, 
in line with those in Vladilo's sample
and consistent with solar. In particular,
$\alpha$ elements (e.g. Si, Ti) are not
found to be overabundant relative to Fe,
and Mn is not significantly underabundant relative to Fe-peak elements.
\footnote{The two Mn/Fe measurements from Vladilo (1998) yield 
[Mn/Fe]$_{\rm corrected} \simeq -0.2$ when adopting 
log(Mn/H)$_{\odot}=-6.47$, from Anders \& Grevese (1989).}

The observed ratio N(Ca\II)/N(Mg\I) in this DLA,
like in two DLA at $z=0.558, 0.5240$ (Vladilo et al.
1997 and references therein), 
 is consistent with 
the predictions for 
 warm gas with the base depletion representative of low density lines
of sight in our Galaxy (Jenkins 1987).
A selection bias would favour lines of sight with a 
base depletion pattern rather than a 
dense depletion, which would be present in gas with a smaller cross-section.

Recent observations of diffuse ISM along several lines of sight
(Meyer et al. 1998 and ref. therein) show evidence for an ISM
abundance of 2/3 solar, rather than solar, 
mainly based on measurements of O and Kr but also consistent with 
N, Zn, S and maybe C. We revised the correction for dust depletion
in the abundance pattern of \heb, assuming intrinsic abundances of 2/3
solar for all elements. The pattern remains almost unchanged, with
deviations lower than 0.05 dex (mostly of $\sim$ 0.02 dex.)

\begin{figure}
\epsfig{figure=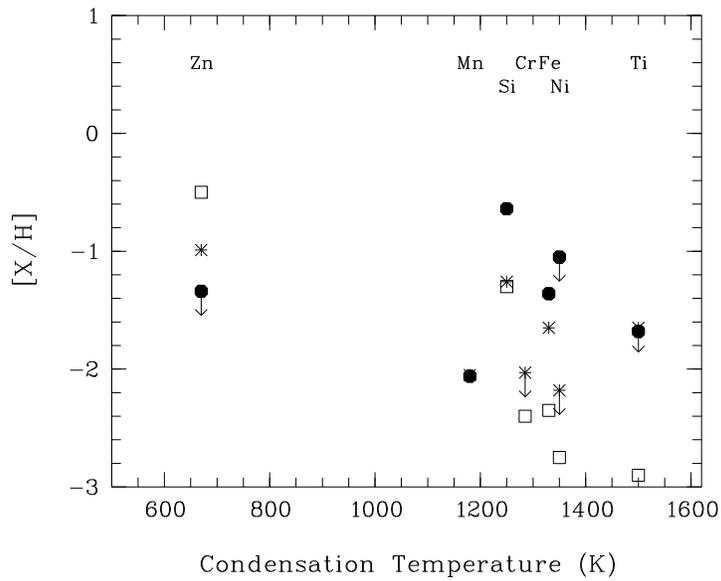,width=15cm}
\caption{\label{temperature}
{Some metal abundances versus condensation temperature for the
$z=0.68$ system (filled circles), the $z=1.15$ system (asterisks) and
in the ISM line of sight towards $\zeta$ Oph from Savage \& Sembach
1996 (open squares).}}

   \end{figure}

The degree to which an element is depleted by dust grains in the
Galactic ISM is known to anticorrelate roughly with 
the temperature at which half the atoms in the
gas phase condense into solid form (e.g. Jenkins 1987).  
A trend suggesting a depletion pattern can
be recognized for this DLA (Fig.~\ref{temperature}), 
but it is mainly based on Zn, the only non-refractory 
element for which abundances could be determined.
 
\subsection{DLA at $z=0.68$  towards HE\,1122$-$1649: no depletion pattern}
The relative abundances observed in the DLA at $z=0.68$ (Table~2)
differ considerably from solar, but follow a
different pattern than the system at $z=1.15$:
\begin{itemize}
\item[--]
The comparable Fe, Ti and Zn content in this system
indicates that the effect of depletion onto dust is negligible,
as expected at this low metallicity. 
Corrections for possible dust depletion  are negligible.
\item[--]
Mn is underabundant relative to Fe, as in the Galactic Halo, and
[Al/Si]$=-0.15 \pm 0.15$ is also compatible with the odd-even effect.
\item[--]
The overabundance of $\alpha$-elements relative to Fe
is observed for Si/Fe and compatible
with the upper limit of O/Fe;
however,  [Ti/Fe]$=-0.3\pm 0.4$.
\end{itemize}
Therefore, the abundance pattern shows similarities with metal-poor Halo stars.
The low depletion degree could indicate
that grain destruction is more efficient than in the Galaxy,
maybe due to increased interstellar shocks from supernova explosions,
as expected in a galaxy with intense star formation. 
Further abundance measurements of other weakly depleted elements (like
S and O) 
would help 
to discern the intrinsic abundance pattern in DLA galaxies.  

Plotting the measured abundances versus condensation temperature is
consistent with the absence of a depletion pattern
(Fig.~\ref{temperature}).  However, as argued by Pettini et
al. (1997a), this lack of correlation should not be overinterpreted at
such a low degree of chemical enrichment.

\vspace{.5cm}

Other lines of sight in the enlarged $z<1.5$ DLA sample where the
observed abundances could be corrected for depletion effects
(Table~\ref{ratios}), also show diverse patterns: e.g., low dust
content is found at $z=0.8598$, $z=1.3726$, and a pattern compatible
with a solar one (or 2/3 of solar) for the corrected abundances at
$z=1.223, 1.149, 1.42, 0.692$.  The different behaviours confirm that
DLA arise in a heterogeneous group, as also inferred from the study of
total abundances.

\subsection{DLA Abundances and the Nucleosynthesis of Nitrogen }

The Nitrogen abundance help us to decipher 
the Galactic chemical enrichment history. At low metallicities, 
N primary production, which takes place
during the asymptotic giant branch phase of intermediate mass stars,
is expected to dominate the secondary N production 
(e.g. Lu et al. 1998 and references therein). The
time-delay between the production of primary N from intermediate mass stars
and elements like O, released from massive stars, would imply a large
scatter in N/O 
at low metal abundances
(e.g. Marconi et al. 1994). However, observations of nearby metal-poor
galaxies do not unambiguously show this pattern.

DLA offer the possibility of studying N abundances for metal-poor
systems tracing early stages of the galactic chemical evolution and,
therefore, of clarifying the N nucleosynthesis.  DLA for which N
abundances --or at least reliable upper limits-- are available, are at
high $z$ (Pettini et al. 1995, Lu et al. 1998,
Centuri\'{o}n et al. 1998 and references therein).
The system at $z=0.68 $ allows the first  study of the N 
content at low redshift.
In spite of the numerous N\I transitions expected in our
spectral range, only upper limits could be derived for N(N\I) due to
the severe blending in the
\ly forest, yielding [N/H]$<-2, $[N/Si]$<-1.36$ and [N/Fe]$<-0.64$.

To compare with other systems, we assume 
that O and the $\alpha$ elements (like S, Si) trace 
each other in DLA, as is the case in the Galaxy 
and in extragalactic H\II regions. 
On the other hand, the lack of depletion onto galactic-like dust 
grains in this system implies that Fe and Si measurements correspond
to the total elemental values, and therefore
[N/O]$\simeq$ [N/S] $\simeq$ [N/Si] is assumed.

Lu et al. (1998) and Centuri\'{o}n et al. (1998) found that the N/Si, N/Fe
and N/S ratios in a sample of DLA at high $z$
show considerable scatter at low metallicities, as predicted
by the standard model  of delayed N primary production. 
The observed N/Fe for our DLA at $z=0.68$ is 
compatible with the range of measurements in Halo stars (cf. Figs.
6 and 7 in Centuri\'{o}n et al.). The low N/Si ($\simeq $ N/O) limit
agrees with the large dispersion and hints 
at a galaxy with recent star formation: O and Si
have recently been released from massive stars and are 
overabundant relative to N,
which has not had time to be primarily
delivered from intermediate mass stars in significant 
amount  and has not been secondarily produced in an 
efficient way at such a low metallicity level.
The $\alpha$/Fe enhancement expected in a 
galaxy with a recent starburst is also observed in this system.
On the contrary, Centuri\'{o}n et al. found no $\alpha$ overabundance in
the few DLA with low N/O values, and suggested that alternative models
are required  to explain those abundance patterns
(e.g. primary N production in massive stars).

\section{Neutral Carbon}
Neutral species should be detectable in gas with a large 
 H\I column density.
The detection of C\I associated with DLA would be particularly
interesting, since the ground-state fine-structure 
excitation depends on the cosmic microwave background radiation,
$T_{CMB}$,
and measuring the relative ground-state populations
one can estimate the excitation temperature and an upper limit for 
$T_{CMB}$ at that epoch (e.g. Ge et al. 1997 
and references therein).

In Table~8 we compare the N(C\I) measured in our systems with other
values from the literature for a large range of $z$, and with
those observed in Galactic lines of sight towards $\zeta$ Oph
(diffuses interstellar medium representative of the Galactic disk) and
$\zeta$ Pup (more tenuous intercloud medium and H\II region).  For the
DLA at $z=1.15$, the absorption features at the expected wavelength of
C\I 1656,1560 yield log$N$(C\I)$=13.68
\pm 0.09$ and 
log$N$(C\I$^*$)$=13.44 \pm 0.07$.
The relative population ratios imply an excitation temperature of 
14.2 K, which is an upper limit to the $T_{CMB}$, since other
mechanisms may contribute 
to the C\I excitation
(Lu et al. 1996). On the other hand,
the complex structure of the $z=1.15$ DLA  and the considerable noise in
the C\I 1560 region makes it difficult to deblend C\I$^*$ from further
C\I components.
For the system at $z=0.68$, we obtained an
upper limit of log$N$(C\I)$<13.45$.  

Comparing these values with the
C\I/H\I ratios known in Galactic gas (Jenkins \& Shaya 1979) we find
that the $z=1.15$ system reproduces the conditions of the Galaxy, as also
found for a DLA at $z=0.8597$ towards PKS\,0454+039 (Boiss\'{e} et
al.\,1998) and $z=1.97$ towards Q\,0013-004 (Ge et al.\,1997).  The
upper limit for the system at $z=0.68$ is still compatible with the
conditions expected for Galactic gas but could also hint at a lower
C\I content as observed for a $z=0.692$ DLA towards 3C\,286
(Boiss\'{e} et al. 1998) and for the DLA at higher $z$ in
Table~8. 

\begin{figure}[hb]\centering   
\epsfig{figure=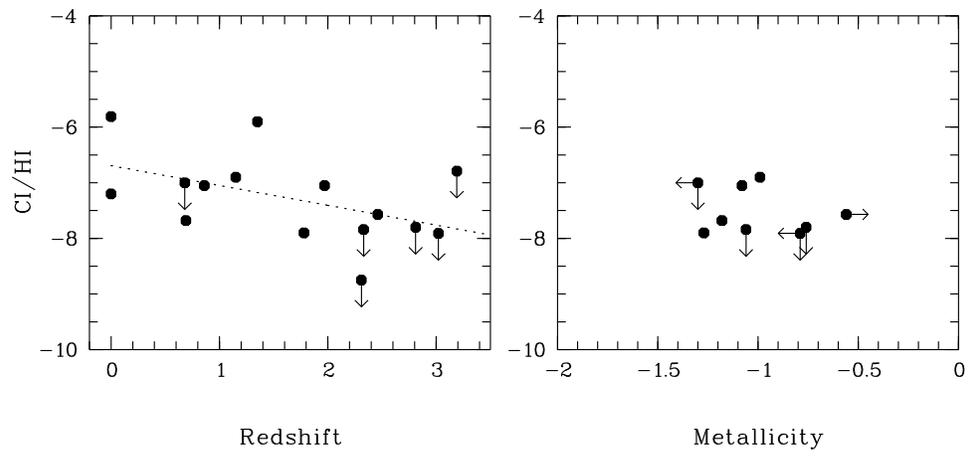,width=17cm}
\caption{\label{cI_evolution}
{Evolution of the C\I/H\I ratio
with redshift (left panel), and with the metallicity (right panel)
References are given in Table~8.}}
\end{figure}

Plotting the C\I/H\I\ ratio versus redshift
(Fig.~\ref{cI_evolution}), a trend of lower C\I/H\I\ with increasing
$z$ can be recognized.  The overplotted linear regression is computed
including upper limits as exact values. Considering that most of high
$z$ measurements are actually upper limits, the observed trend would
be more pronounced.  The evolution of the metagalactic ionizing
background (more intense at high $z$ than at $z\simlt 1$, e.g. Haardt
\& Madau 1996) is probably the main reason of the C\I/H\I evolution
with $z$.  On the other hand, the local presence of dust could also be
playing a role, since dust grains could shield more efficiently the
gas from being photoionized.  In fact, the DLA in Table~8
with a low dust content show a low C\I/H\I ratio,
regardless of the redshift.

\section{Molecules in DLA}

We do not see (H$_2$) or CO absorption, consistently with the low 
metal abundances.

Molecular hydrogen (H$_2$) has numerous transitions at $\lambda <
1100$ \AA\ and carbon monoxide (CO) at $\lambda < 1550$ \AA\ .
Because of the blending problem it is better to look for
absorption-free windows where strong H$_2$ is expected than looking
for strong lines themselves (Ge \& Bechtold 1997). We do not find
absorption at the expected positions of some strong H$_2$ lines
(e.g. around $\lambda_{rest} = 1094, 1079$) associated with the DLA at
$z=0.68$.  
The H$_2$/H ratio
is small compared to Galactic interstellar clouds,
a common result for DLA (Ge \&
Bechtold, 1997).  For the DLA at $z=1.15$,
the spectra do not cover 
the region where H$_2$ absorption is expected.

We searched for strong CO transitions ($\lambda_{rest}$ = 1477.568,
1477.355, 1509.75, 1419.046, 1392.525, 1544.451 \AA)
and obtained $2\sigma$ upper limits of 
logN(CO)$<14.4$ for the DLA at $z=0.68$ and 
logN(CO)$<13.8$ at $z=1.15$ (where only
$\lambda_{rest}$ = 1544 \AA\ falls in the available wavelength range).
Contrasting these values with the upper limits derived by Lu et al. (1998)
at the Keck HIRES resolution for several DLA at high $z$, we find that
our CO limits are  roughly comparable to the values deduced for
diffuse interstellar clouds in the Milky Way.

\section{Kinematics: Complex Structure at Moderate \boldmath $z$ \unboldmath}

The DLA at $z=0.68$ towards \hea extends over $\sim 230$ \km
in velocity space
plus a satellite component at $-100$ \km.
As we can see in Figs.~\ref{ions_1122A} and \ref{ions_1122B}, the profiles of low ions observed 
at high resolution
(cf. Fe\II, Mg\II, Mn\II) track closely one another  and show an
edge-leading asymmetry, especially if we consider the 
feature from $-20$ to 100 \km.  

Remarkably, the Mg\I 2852 profile shows a different asymmetry: the
main feature lies between 100 and 150 \km, while Mg\II -- along
with other singly ionized ions -- shows the strongest 
absorption between $\sim -20$ and 100 \km. This suggests that the
region between 100 and 150 \km is less ionized, since the feature
could not be identified as a metal line associated with another
absorber.  
The difference between Mg\I\ and Mg\II\ profiles could hint at a
non-negligible contribution of an intervening H\II\ region for this
system.  The ionized gas would produce an extra contribution to the
column densities of species with an ionization potential larger than
13.6 eV, so that the measured abundances of ions like Fe\II, Mn\II,
Si\II and Zn\II\ would correspond to both regions. The 
metal abundances associated with the H\I\ system would therefore be lower.

The low-ion profiles of the system at $z=1.15$ towards \heb 
(Figs.~\ref{ions_0515A} and \ref{ions_0515B}) span more than 700 \km, which is the largest extent in the 
velocity space
seen for a DLA.  As pointed out before, three main structures can be
recognized: from approximately $-575$ to $-270$ \km, from $-260$ to
$-100$ \km and the main absorber from $-80$ to 140 \km.  This main
complex
shows no ordered profile and is consistent with absorber components
distributed randomly in $v$.

The remarkably complex structure 
of low and intermediate $z$ DLA provide new hints about their nature.
Two kinds of models have been discussed to explain kinematic
characteristics of DLA at high $z$: (1)
thick rotating disks evolving without merging
to form present-day galaxies (Prochaska \& Wolfe 1997a,b,1998),
and (2) merging protogalactic clumps in the 
standard Cold Dark Matter cosmology (Haehnelt et al. 1998, McDonald \&
Miralda-Escud\'{e} submitted).
The model of rotating thick galactic disks
can explain the edge-leading asymmetry of profile structures extending
up to $\sim 250$ \km. 
An asymmetric structure where the absorption decreases monotonously 
is found in the system at $z=0.68$, but the one at $z=1.5$
shows a non-ordered profile.  The large extent of
the systems suggests that they cannot be single rotating
disks.

Alternatively, several authors (e.g. Matteuci et al. 1997, Vladilo et
 al. 1997) have suggested that dwarf galaxies rather than spiral
 galaxies could be responsible for DLA.  The extent of the kinematic
 profiles of our two new DLA are again too large  to be
 explained by individual dwarf galaxies.
On the other hand,
Khersonsky \& Turnshek (1996) showed that DLA can be associated with
neutral gas in giant hydrogenous clouds that could be associated
with any type of galaxy or protogalaxy.

Considering only the deepest part of the absorption ($\tau \sim 0.7$) most 
profiles in a sample of high $z$ DLA studied by Ledoux et al. (1998)
show $\Delta v < 150$ \km and a trend of decreasing
velocity 
with increasing redshift. Under the
same assumptions, our DLA at $z=0.68$ and $1.15$ 
match fairly well this behaviour, which is expected in CDM 
structure formation scenario (e.g. Kauffmann 1996),
but not if DLA are disks which do not evolve kinematically.
Similarly to the models
described by Haehnelt et al. 1998 for high $z$, 
the velocity broadening in the clumps at intermediate $z$ 
would be due to a mixture 
of random motions, infall, merging
and also rotation, which would still be important at smaller scales 
and could be responsible for the edge-leading 
asymmetry observed in some of the clumps.

Wolfe \& Prochaska (1998) found a correlation between kinematics and
metallicity for a sample of 17 DLA. 
At $z=1.6-3$ 
they  detected no systems like our two, with a large extent
in the velocity space ($\Delta v >120$ \km) and
low metallicity ([Zn/H] $\leq -1$) at the same time.
On the other hand, the new systems 
do show moderate \nh\ of  $\simeq 10^{20.4}$ cm$^{-2}$, following the trend 
of larger $\Delta v$ to lower \nh observed 
for high redshift DLA (Wolfe \& Prochaska 1998).

The ions associated with the DLA at $z=1.15$ towards \heb seem to
track each other fairly well, but the lack of unsaturated lines from
several ions at some velocities
makes it difficult to establish the variation of the relative
abundances. 
However, the strongest feature in the profiles of unsaturated Fe\II
and Mn\II show two absorption features of comparable intensity, 
while for Zn\II and Si\II, the first one is clearly larger. 
Species weakly related to dust in the Galactic ISM
seem to show a different profile than those
corresponding to refractory elements.  
This could mean that there is more dust present in the components between
$-50$ and $+50$ \km, possibly associated with a higher metallicity
level, and this dust could be partly concealing the edge-leading
asymmetry.

\section{Associated Highly-Ionized Gas.}

Damped Ly$\alpha $ systems studied at high $z$ often show considerable
Si\IV and C\IV absorption with profiles resembling each other, but usually 
different from low ions (e.g. Lu et al. 1996).  This suggests that
highly-ionized gas is produced in distinct physical regions, which
could be surrounding and shielding the low-ionized gas.
In our two DLA at moderate redshift, we do not
find strong lines from
highly-ionized species  covering
a much larger extent than the weakly-ionized species.

\begin{itemize}
\item[--]{\em DLA at $z=0.68$ towards HE\,1122$-$1649:}
 Fig.~\ref{ions_1122B} (right panel) shows
absorption near the expected positions of C\IV, O\VI,
 Si\III and Si\IV, none of which is unambiguously present.

\item[--]{\em DLA at $z=1.15$ towards HE\,0515$-$4414:} 
Only weak C\IV absorption with $N$(C\IV) $= 13.18 \pm 0.12$ 
was found associated with the main complex at $z=1.15$ (from $-80$ to 140
\km), contrary to most of DLA, but like
$z=0.692$ towards 3C286 (Boiss\'{e} et al. 1998), where no associated high
ions could be found except for weak C\IV. This suggests that the bulk
of gas is weakly ionized and shielded from the ionizing radiation
field.  
Stronger absorption features for the C\IV doublet yielding
N(C\IV)= 14.38 $\pm 0.17$ could be fitted in the noisy region
extending from $\sim -550$ to $-200$ \km (Fig.~\ref{ions_0515A}), which 
is less prominent in the
low-ion profiles. 
\end{itemize}
Al\III\ measured in the DLA follows the low ionization species (Al\III\
is observed for the three main substructures at $z=1.15$), but it is
present in a much lower amount than Al\II. These results confirm the
very low ionization degree in these systems and suggest that Al\III\,
is probably produced in the same physical region as the low ionization
species. 

The absence of highly ionized gas associated with the large extents of
low ionized regions suggests that the ionizing background is weaker
than at high $z$, as expected from the cosmological expansion and from
models of the ionizing
background (e.g. Haardt \& Madau 1996).

\section{Summary}

We have studied the properties of the DLA at $z=0.68$
towards \hea and  at $z=1.15$ 
towards \heb. A further DLA candidate at low redshift ($z=0.28$)
has been found in the line of sight of \heb.
Very few DLA at $z<1.1$ had been analyzed up to now
and enlarging the sample is essential  to understand the
nature of DLA, cosmic chemical evolution and  possible 
selection effects which could be biasing our view of the Universe.
The main results of our analysis, interpreted in comparison
with DLA at other $z$, are summarized in this section:
\begin{itemize}
\item A large number of associated metal lines have been detected at high
resolution.  Metal abundances are found to be rather low (of 1/10 and
$<1/20$ of solar value)
and comparable to 
high-$z$ DLA. Therefore, the new values confirm the large scatter at
any $z$, rather than   
a clear increase in the metallicity with decreasing $z$
for the DLA population. 
The new systems also confirm the heterogeneity of the DLA population
(in agreement with the results from imaging studies, e.g. by Le Brun et
al. 1997).
\item 
The relative abundance pattern for the DLA at $z=1.15$ in the line of
sight of the very bright
\heb indicates a significant degree of 
depletion, with a dust-to-metal ratio comparable to the Galactic
value.  The intrinsic pattern, after correcting for dust effects, is
comparable to solar.
\item The system at $z=0.68$ towards
\hea\ presents a negligible dust content and
its abundance pattern resembles that in metal-poor Halo stars. 
Relative abundances suggest that most metals in the $z=0.68$
DLA where made recently.
Primary N production in massive stars is not required to explain
the abundance pattern.
\item The search for neutral carbon shows that systems
without significant amounts of dust are not
as shielded from ionizing sources as systems where dust is present.
C\I/H\I\ increases with decreasing $z$, as expected
from the evolution of the metagalactic ionizing background.
\item The non-detection of H$_2$ or CO molecules is congruent with
the low metal abundances.
\item The high-resolution data reveal a very complex structure for the
low-ion profiles. The system at $z=0.68$ spans over $\sim$ 300 \km and
shows an edge-leading asymmetry. The $z=1.15$ system shows at least
three main substructures over more than 700 \km.
This suggests that the line of
sight intersects a group of 
galaxies or subgalactic units.
The velocity broadening could be due to a mixture of random
motions, infall, merging and rotation.
\item Only weak absorption from highly-ionized species were
detected and it probably originates in regions
distinct from the low-ionized gas.
\end{itemize}

Further surveys will provide new DLA candidates, which should be
observed with the new echelle spectrographs and, at moderate redshift,
also with imaging techniques in order to detect their galactic
counterparts.  Enlarging the DLA sample and clarifying their nature
will help us to understand the cosmic chemical evolution and to
discern among models of galaxy formation.

\newpage
\tablecaption{Line fits in the DLA at $z=0.68$ towards \hea \label{tbl-1}}
\tablehead{\hline}
\tabletail{\hline}
\begin{supertabular}{p{.5cm}p{.5cm}p{1.2cm}rp{.4cm}rp{.6cm}}

\scriptsize
& $Ion$ & $z$
& log\,$N$
& $\sigma_{N}$
& \hspace{-1cm} $b$ 
& $\sigma_b$
\\ \hline

Al\I  & 3083 & 0.6819088 & $12.17$ & 0.07 & 10.0  & 0.0  \\
Al\II & 1670 & 0.6820851 & $14.14$ & 0.11  & 30.0  & 0.0  \\
Al\III & 1862 & 0.6822000 & $<13.00$ &   & 30.0  & 0.0  \\
C\I & 1656 & 0.6821087 & $<13.45$ &    & 30.0  & 0.0   \\
C\II & 1334 & 0.6822630 & $18.30$ & 0.29   & 30.0  & 0.0   \\
Fe\I & 2523 & 0.6818900 & $<11.44$&  & &\\
Fe\II & 2600 & 0.6813161 & 12.44 & 0.02 & 3.00 & 0.09 \\
      & 2374 & 0.6818802 & 14.05 & 0.04 & 4.68 & 0.64 \\      
      & 2374& 0.6819466 &  13.52 &   0.13 &   4.48  &  1.99  \\
      & 2374 & 0.6820377 &  13.71&    0.07 &   8.71 &   2.10 \\
    &2374& 0.6821409&   12.98  &  0.07 &   4.32  &  4.56 \\
    &2374& 0.6822150&   13.69 &   0.13&   11.25  &  3.75  \\
   &2374&  0.6823508 &  13.52  &  0.08 &   9.35 &   1.79  \\
    &2374& 0.6824777 &  12.72 &   0.10 &   3.00 &   0.00  \\
    &2374& 0.6825430 &  13.60 &   0.04 &   3.00 &   0.00  \\
    &2374& 0.6825936 &  13.32 &   0.05 &   3.00 &   0.00 \\
    &2374& 0.6826816 &  13.16 &   0.08&   10.32 &   2.69  \\
   &2374&  0.6828596 &  12.65 &   0.10 &   3.00 &   0.00  \\
   &2374&  0.6829300 &  13.06&    0.08 &   3.00  &  0.00   \\
    &2374& 0.6830164 &  13.05 &   0.13 &  10.57 &   5.00  \\
Mg\I & 2852 &      0.6819096 &  11.37 &   0.13&    3.15  &  0.48 \\
     & 2852 &      0.6821620 &  11.62 &   0.04&   22.98   & 3.81 \\
     & 2852 &      0.6825483 &  11.49 &   0.43&    2.59   & 1.05 \\
     & 2852 &      0.6826613 &  11.32 &   0.11&    4.28   & 1.73 \\
     & 2852 &      0.6829651 &  11.74 &   0.14&    2.39   & 0.19 \\
     & 2852 &      0.6832639 &  11.50 &   0.17&    1.96   & 0.34  \\
Mg\II & 2796 &     0.6813161 &  12.74 &   0.04 &   4.43  &  0.33  \\
      & 2803 &     0.6813162  & 12.74 &   0.04 &   4.43 &   0.33  \\
      & 2803 &      0.6818795 &  14.42 &  0.37 & 2.81  & 0.24   \\
      & 2796 &      0.6818794 &  14.42 &  0.37 & 2.81  & 0.24  \\
      & 2803 &      0.6819472 &  13.21 &  0.04 & 14.62  & 0.56 \\
      & 2796 &      0.6819472 &  13.21 &  0.04 & 14.62  & 0.56  \\
      & 2803 &      0.6820363 &  13.52 &  0.05 & 9.25  & 0.89 \\
      & 2796 &      0.6820364 &  13.52 &  0.05 & 9.25  & 0.89  \\
      & 2803 &      0.6821541 &  13.40 &  0.17 & 3.70  & 0.94  \\
      & 2796 &      0.6821541 &  13.40 &  0.17 & 3.70  & 0.94 \\
      & 2803 &      0.6822218 &  13.79 &  0.18 & 5.57  & 0.82  \\
      & 2796 &      0.6822218 &  13.79 &  0.18 & 5.57  & 0.82 \\
      & 2803 &      0.6823556 &  13.43 &  0.02 & 10.32  & 0.75  \\
      & 2796 &      0.6823556 &  13.43 &  0.02 & 10.32  & 0.75  \\
      & 2803 &      0.6824673 &  12.84 &  0.11 & 3.50  & 0.62  \\
      & 2796 &      0.6824672 &  12.84 &  0.11 & 3.50  & 0.62  \\
      & 2803 &      0.6825500 &  13.64 &  0.27 & 3.96  & 0.48  \\
      & 2796 &      0.6825500 &  13.64 &  0.27 & 3.96  & 0.48 \\
      & 2803 &      0.6826048 &  13.23 &  0.73 & 2.18  & 1.45  \\
      & 2796 &      0.6826049 &  13.23 &  0.73 & 2.18  & 1.45 \\
      & 2803 &      0.6826891 &  13.22 &  0.01 & 12.23  & 0.55 \\
      & 2796 &      0.6826891 &  13.22 &  0.01 & 12.23 & 0.55 \\
      & 2803 &      0.6828574 &  11.27 &  0.47 & 13.08  & 7.72 \\
      & 2796 &      0.6828575 &  11.27 &  0.47 & 13.08 & 7.72 \\
      & 2803  &    0.6829114 &  12.58   &  0.13  &  7.16 &   0.91 \\
      & 2796 &     0.6829112 &  12.58 &   0.13 &   7.16 &0.91 \\
      & 2803   &   0.6829443 &  12.25 & 0.27  &  2.30  &  0.71 \\
      & 2796   &   0.6829442 &  12.25  &  0.27 & 2.30 & 0.71 \\
Mn\I & 2799 & 0.6818900 & $< 11.67$ &   &   &    \\
Mn\II & 2576 & 0.6818700 & $ 11.68 $ & 0.09  &  2.93 & 2.57 \\
      & 2594 & 0.6818699 & $ 11.68 $ & 0.09  &  2.93 & 2.57 \\
      & 2576 & 0.6819273 & $ 11.55 $ & 0.11  &  1.55 & 2.65 \\
      & 2594 & 0.6819273 & $ 11.55 $ & 0.11  &  1.55 & 2.65 \\
N\I & 1199 &  0.6818900 & $ <14.5 $ &   & 30.0& 0.00   \\
N\V & 1238 &  0.6818900 & $ <14.0 $ &   &  &  \\
    & 1242 & & & & & \\
Ni\I & 2346 &  0.6818900 & $ <12.07 $ &   & &    \\
Ni\II & 1317 &  0.6818900 & $ <13.65 $ &   & &   \\
O\I & 1039 &  0.681890 &  $<17.00$ &    &  30.0 &   0.00 \\
O\VI & 1031 &       0.6821704 &  18.94  &  0.09 &   0.00 &  24.53  \\
S\I & 1807 & 0.6820000 &$ < 14.05 $     & & &  \\
S\III & 1012 & 0.6820000 &$ < 14.99 $     & &  & \\
Si\I & 2515 & 0.6818900 &$ < 14.58 $     & & & \\
Si\II & 1193 & 0.6819438 &  15.32  &   0.09  &  25.00  &   0.00 \\
      & 1190 & 0.6819438 &  15.32  &   0.09 &   25.00  &   0.00  \\
      & 1193 & 0.6825498  & 14.36   &  0.12  &  16.00  &   0.00 \\
      & 1190 & 0.6825498  & 14.36   &  0.12  &  16.00  &   0.00  \\
Si\IV & 1402   &   0.6823700 &  14.32 &   0.23 &  30.  & 0.0\\
Ti\II & 3384   &   0.6818927 &  11.70 &   0.28 &  1.19  & 0.7\\
Zn\I  & 2139 & 0.68189 &$ < 11.28 $     & & &  \\
Zn\II & 2026 & 0.68189 &$ < 11.76 $     & & &  \\

\end{supertabular}

\vspace{2.cm}


\tablecaption{Metal abundances of the DLA at $z=0.68$ with 
log\nh = $20.45 \pm 0.15$. 
We adopt the classical definition 
\mbox{ [X/Y] $\equiv \log$ (X/Y)$_{\rm obs} - \log$ (X/Y)$_{\odot}$}, 
where solar values 
are taken from Anders \& Grevesse (1989). 
\label{tbl-2}}
\tablehead{\hline}
\tabletail{\hline}
\begin{supertabular}{ccccc} 

Element & log\,$N$(cm$^{-2}$) & Total Abundance
& [Metal/Fe] & [Metal/Zn]\\ \hline

Al & $14.14\pm 0.11$ & $-0.79 \pm 0.12$ & $0.57 $ & $>0.55 $ \\
C  & $18.30\pm 0.29$ & $1.28 \pm 0.29$ & $2.64$& $>2.62 $\\
Fe &$14.60\pm 0.28$ & $-1.36 \pm 0.28$ & & $>-0.02 $\\
Mg & $>14.71$ &$> -1.32$      & $>0.04$&\\
Mn &$11.92\pm 0.14$ & $-2.06 \pm 0.15$ & $-0.7 $& $>-1.02 $\\
N  & $<14.5$ &$< -2. $  &$<-0.64$& \\
Ni &$<13.65$ & $< -1.05 $ &$<0.31$&\\
O  &$<17 $ & $<-0.38 $ & $<0.98 $ & $ $\\
Si &$15.36\pm 0.12$ & $-0.64\pm 0.13$ & $ 0.72$   & \\
Ti &$11.70\pm 0.28$ &  $-1.68 \pm 0.28$ &$-0.32 $& $>-0.34 $\\
Zn & $<11.76$ & $< -1.34$ & $<0.02$&\\

\end{supertabular}

\newpage


\tablecaption{Line fits in the DLA at $z=1.15$.\label{tbl-3}}
\tablehead{\hline}
\begin{supertabular}{p{.5cm}p{.5cm}p{1.2cm}rp{.4cm}rp{.6cm}}
\scriptsize
& $Ion$ & $z$
& log\,$N$
& $\sigma_{N}$
& $b$ 
& $\sigma_b$ \\\hline

 Al\II & 1670 & 1.1468446 & 11.44 & 0.19 &  7.03 & 5.34 \\
      &      & 1.1469733 & 12.40 & 0.20 &  2.77 & 0.28 \\
      &	     & 1.1470984 & 11.80 & 0.20 &  2.37 & 8.48 \\
      &      & 1.1472185 & 12.08 & 0.19 &  8.80 & 6.53 \\ 
      &      & 1.1475825 & 12.01 & 0.06 &  1.43 & 0.06 \\
      &      & 1.1480030 & 11.36 & 0.10 &  9.38 & 3.58 \\
      &      & 1.1484466 & 11.98 & 0.08 &  1.26 & 0.07 \\
      &      & 1.1487808 & 12.53 & 0.45 &  0.21 & 2.52 \\
      &      & 1.1490958 & 13.02 & 0.38 &  2.85 & 0.35 \\ 
      &      & 1.1495245 & 11.80 & 0.08 &  7.67 & 2.55 \\
      &      & 1.1498086 & 12.69 & 0.08 &  1.07 & 0.23 \\
      &      & 1.1507328 & 12.13 & 1.09 & 23.51 & 22.7 \\ 
      &      & 1.1508443 & 15.42 & 0.87 &  7.71 & 1.71 \\
      &      & 1.1511260 & 12.64 & 0.15 &  9.91 & 3.84 \\ 
      &      & 1.1512798 & 13.53 & 0.82 &  5.13 & 2.02 \\ 
      &      & 1.1513751 & 12.05 & 0.77 & 25.35 &35.08 \\ 
						      
Al\III & 1854& 1.1490991 & 11.88 & 0.06 &  2.84 & 1.30 \\ 
       & 1862&           &       &      &       &      \\ 
       & 1854& 1.1508672 & 12.07 & 0.07 & 14.73 & 3.17 \\ 
C\I   & 1656 & 1.1508082 & 13.56 & 0.04 & 9.31  & 1.43 \\
      & 1560 & 1.1508083 &  &  &   &   \\ 
      & 1656 & 1.1513586 & 13.05 & 0.09 & 12.68  & 1.03 \\
      & 1560 & 1.1513586 &  &  &   &  \\ 
C\I$^*$   & 1656.2 & 1.1507983 & 13.44 & 0.07 & 7.48  & 2.45  \\
C\I$^*$   & 1657.9 & 1.1507983 &  &  &   & \\

 C\IV& 1548 &      1.1471568 &  13.63 &   0.06&   20.69&    3.11 \\
     & 1550 &      1.1471567 &  13.63 &   0.06&   20.69&    3.11 \\
     & 1548 &      1.1475769 &  13.55 &   0.08&    9.33&    1.56 \\
     & 1550 &      1.1475769 &  13.55 &   0.08&    9.33&    1.56 \\
     & 1548 &      1.1480157 &  13.68 &   0.08&    7.03&    1.33 \\
     & 1550 &      1.1480159 &  13.68 &   0.08&    7.03&    1.33 \\
     & 1548 &      1.1482468 &  13.74 &   0.07&   10.06&    2.00 \\
     & 1550 &      1.1482468 &  13.74 &   0.07&   10.06&    2.00 \\
     & 1548 &      1.1485275 &  13.61 &   0.08&   24.69&    5.03 \\
     & 1550 &      1.1485275 &  13.61 &   0.08&   24.69&    5.03 \\
     & 1548 &      1.1491662 &  13.26 &   0.09&  11.06 &    3.27 \\
     & 1550 &      1.1491662 &  13.26 &   0.09&   11.06&    3.27 \\

C\IV  & 1548 & 1.1508356 & 13.18 & 0.12 & 10.12 & 4.35  \\
      & 1550 &           &       &      &       &       \\

Ca\II & 3934 & 1.1508502 & 11.87 & 0.23 & 7.23  & 4.79  \\
      &      & 1.1509476 & 12.17 & 0.14 & 30.14 & 8.88  \\
      &      & 1.1513215 & 12.25 & 0.05 & 12.60 & 1.82  \\
Cr\II & 2056 & 1.150800  &$<12.10$&     &       &     \\     
Fe\I  & 2484 & 1.150800& $<11.57$& & & \\

Fe\II & 2600 & 1.1469649 & 12.43 & 0.05 &  2.30 & 0.51  \\
      & & 1.1469650 & 12.43 & 0.05 &  2.30 & 0.51  \\   
      & & 1.1471601 & 11.93 & 0.15 & 10.52 & 2.81  \\   
      & & 1.1471599 & 11.93 & 0.15 & 10.52 & 2.81  \\   
      & & 1.1472481 & 12.66 & 0.05 & 50.32 & 5.67  \\   
      & & 1.1472480 & 12.66 & 0.05 & 50.32 & 5.67  \\   
      & & 1.1479497 & 12.58 & 0.03 & 21.17 & 1.63  \\   
      & & 1.1479497 & 12.58 & 0.03 & 21.17 & 1.63  \\   
      & & 1.1490909 & 13.11 & 0.13 &  2.40 & 0.21  \\
      & & 1.1490909 & 13.11 & 0.13 &  2.40 & 0.21  \\
      & & 1.1495110 & 12.54 & 0.02 &  8.42 & 0.78  \\
      & & 1.1495110 & 12.54 & 0.02 &  8.42 & 0.78  \\
      & & 1.1498045 & 12.21 & 0.06 &  4.82 & 3.59  \\
      & & 1.1498046 & 12.21 & 0.06 &  4.82 & 3.59  \\
      & & 1.1499078 & 12.06 & 0.10 &  2.89 & 1.22  \\
      & & 1.1499077 & 12.06 & 0.10 &  2.89 & 1.22  \\
  & 2374& 1.1508005 &  13.75  &  0.01 &   9.54 &   0.64 \\
  & &     1.1509480 &  13.68  &  0.05 &   1.73 &   0.12 \\
  & &     1.1510571 &  13.13  &  0.06 &   2.30 &   0.62 \\
  & &     1.1512597 &  13.64  &  0.03 &  15.35 &   1.27 \\
  & &     1.1512957 &  13.18  &  0.18 &   1.01 &   6.68 \\
Mg\I  & 1827 & 1.1508058 & 12.86 & 0.09 & 16.34 & 4.68 \\
      & 2026 &           &       &      &       &      \\
      & 2852 & 1.1508093 & 12.06 & 0.02 &  7.19 & 0.52 \\
      &      & 1.1509610 & 12.29 & 0.02 & 29.61 & 1.07 \\
      &      & 1.1513028 & 11.95 & 0.02 & 11.32 & 0.54 \\
Mg\II & 2796 & 1.1469659 & 12.79 & 0.23 &  2.88 & 1.04 \\
      & 2803 & 1.1469659 & 12.79 & 0.23 &  2.88 & 1.04 \\
      & 2796 & 1.1470861 & 13.00 & 0.05 & 28.34 & 2.83 \\
      & 2803 & 1.1470863 & 13.00 & 0.05 & 28.34 & 2.83 \\
      & 2796 & 1.1471353 & 12.34 & 0.08 &  4.53 & 2.01 \\
      & 2803 & 1.1471351 & 12.34 & 0.08 &  4.53 & 2.01 \\
      & 2796 & 1.1472504 & 12.29 & 0.16 &  5.40 & 2.42  \\
      & 2803 & 1.1472507 & 12.29 & 0.16 &  5.40 & 2.42  \\
      & 2796 & 1.1474081 & 12.11 & 0.11 &  5.15 & 1.77  \\
      & 2803 & 1.1474082 & 12.11 & 0.11 &  5.15 & 1.77  \\
      & 2796 & 1.1475828 & 12.25 & 0.02 &  7.17 & 0.60  \\
      & 2803 & 1.1475830 & 12.25 & 0.02 &  7.17 & 0.60  \\
      & 2796 & 1.1479594 & 12.62 & 0.01 & 26.94 & 0.91  \\
      & 2803 & 1.1479595 & 12.62 & 0.01 & 26.94 & 0.91  \\
      & 2796 & 1.1485009 & 12.34 & 0.02 & 20.98 & 1.22  \\
      & 2803 & 1.1485010 & 12.34 & 0.02 & 20.98 & 1.22  \\
      & 2796 & 1.1491084 & 12.85 & 0.01 &  9.75 & 0.19  \\ 
      & 2796 & 1.1495093 & 12.34 & 0.03 &  7.06 & 0.63  \\ 
      & 2796 & 1.1496212 & 12.43 & 0.05 & 34.58 & 5.65  \\ 
      & 2796 & 1.1498080 & 12.77 & 0.55 &  2.10 & 0.51  \\ 
      & 2796 & 1.1499032 & 12.21 & 0.08 &  6.85 & 1.32  \\ 
      & 2803 & 1.1503466 & 11.65 & 0.16 &  6.43 & 3.15  \\
      & 2803 & 1.1506114 & 12.79 & 0.11 & 21.11 & 3.89  \\
      & 2803 & 1.1508448 & 13.80 & 0.04 & 15.89 & 0.87  \\
      & 2803 & 1.1509888 & 13.41 & 0.12 & 23.32 & 4.36  \\
      & 2803 & 1.1512514 & 13.41 & 0.08 & 31.96 & 1.92  \\
      & 2803 & 1.1512749 & 15.11 & 0.42 &  4.94 & 0.49  \\
      & 2803 & 1.1517754 & 11.74 & 0.11 &  9.73 & 4.27  \\
Mn\II & 2576 & 1.1507992 & 11.67 & 0.08 &  4.61 & 2.30  \\
      & 2594 & 1.1507992 & 11.67 & 0.08 &  4.61 & 2.30  \\
      & 2606 & 1.1507992 & 11.67 & 0.08 &  4.61 & 2.30  \\
Mn\II & 2576 & 1.1512835 & 11.58 & 0.11 &  4.09 & 5.22  \\ 
Na\I  & 3303 & 1.1508000 & $<12.9$&  &  &    \\
Ni\II & 1741 & 1.1507868 & $<12.52$ & 0.12 &  5.19 & 3.94 \\
S\I   & 1807 & 1.1508000 & $< 12.52$&     &       &          \\
Si\II & 1808 & 1.1507919 & 14.51 & 0.05 &  7.2 & 2.8 \\
Si\II & 1526 &           &       &      &       &     \\
Si\II & 1808 & 1.1509281 & 13.98 & 0.14 & 2.08 & 4.19 \\
Si\II & 1526 &           &       &      &       &     \\
Si\II & 1808 & 1.1512949 & 14.12 & 0.11 & 2.49 & 3.19 \\ 
Si\II & 1526 &           &       &      &       &     \\
Zn\II & 2026 & 1.1513157 & 12.11 & 0.04 & 9.15 & 1.66 \\
\end{supertabular}
\vspace{1.cm}


\vspace{1.cm}

\tablecaption{Metal column densities and total 
abundances for the DLA at $z=1.15$ adopting
log\nh = $20.45$.
Feature I, II and III correspond to the substructures from  $-575$ to
$-270$, from $-260$ to $-100$ and from $-80$ to 140 \km, respectively.
\label{tbl-4}}
\tablehead{\hline}
\tabletail{\hline}
\begin{supertabular}{lccc}

\small
$Total Abundance$  &
$N_{Feature I}$ &
$N_{Feature II}$&
$N_{Feature III}$ \\ \hline
\small
$ \mbox{[Al/H]} \ge 0.5$ & $ 13.01 \pm 0.61 $ & $13.20 \pm 0.40$ & $\ge 15.4$  \\
$ \mbox{[Ca/H]} > -2.19 $ & & & $12.60 \pm 0.27 $ \\
$ \mbox{[Cr/H]} \le -2.03$  & & &  \\
$ \mbox{[Fe/H]}=-1.65$&$13.08\pm 0.17$&$13.28\pm 0.17$&$14.24\pm 0.20$\\
$ \mbox{[Mg/H]}\ge -0.8$ & $ 13.47 \pm 0.22 $ & $13.29 \pm 0.56$& $\ge 15.15$\\
$ \mbox{[Mn/H]} = -2.05 $ &  & $< 11.4$ & $11.93 \pm 0.11$\\
$ \mbox{[Ni/H]}< -2.18 $ & & & $<12.52 $  \\
$ \mbox{[Si/H]}=-1.26 $ &&$<13.8 $ &$14.74 \pm 0.18$  \\
$ \mbox{[Ti/H]} \le -1.65$  & & & $< 11.7$  \\
$ \mbox{[Zn/H]}=-0.99 $ & &$<11.5$ &$12.11 \pm 0.04$  \\
\end{supertabular}

\newpage

\tablecaption{ Relative Metal abundances for the DLA at $z=1.15$. \label{tbl-5}}
\tablehead{\hline}
\tabletail{\hline}
\begin{supertabular}{lccc}
\small
$Element X$  &
$[X/Fe]_{Total}$ &
$[X/Fe]_{Main}$ &
$[X/Zn]_{Total}$ \\ \hline
\small
Al   & $>2.12$  & $>2.19$    & $>1.46$  \\
Ca\II& $-0.54$  & $-0.47$    & $-1.2$   \\
Cr   & $<-0.38$ & $<-0.31$   & $<-1.04$ \\
Fe   &          &            & $-0.66$  \\
Mn   & $-0.40$  & $-0.59$    & $-1.06$  \\
Ni   & $<-0.53$ & $<-0.46$   & $<-1.19$ \\
Si   & $ 0.39$  & $ 0.46$    & $-0.27$  \\
Ti   & $<-0.03$ & $<-0.07$   & $<-0.66$ \\
Zn   & $0.66$   & $0.73$     &          \\

\end{supertabular}


\vspace{2.cm}

\tablecaption{ Line parameters for the absorber at $z=0.28$ towards
\heb\label{tbl-6}}
\tablehead{\hline}
\tabletail{\hline}
\begin{supertabular}{cccccccc}
\small
&$Ion$  
& $z$
&log\,$N$(cm$^{-2}$)
&$\sigma_{N}$)
&$b$\,(km~s$^{-1}$)
&$\sigma_b$ \\\hline
\small
Al\I  & 3969 & 0.2813500  & $< 11.91 $   &       &      \\
Ca\II & 3969 & 0.2813500  & $< 11.57 $   &       &           \\ 
Fe\I  & 3021 & 0.2813500  & $< 12.12 $   &       &            \\
Fe\II & 2600 & 0.2813753 & 13.14 & 0.09 &15.00 & 0.00 \\
      & 2600 & 0.2817716 & 12.96 & 0.20 & 5.29  & 4.82 \\
Mn\II & 2576 & 0.2813256 & 12.46 & 0.15 & 8.00 & 0.00 \\
      & 2606 & 0.2813256 & 12.46 & 0.15 & 8.00 & 0.00 \\
Mg\II & 2796 & 0.2812490 & 12.44 & 0.08 & 13.93 & 3.54 \\
      & 2803 & 0.2812490 & 12.44 & 0.08 & 13.93 & 3.54 \\
      & 2796 & 0.2813520 & 12.77 & 0.27 &  6.92 & 2.42 \\
      & 2803 & 0.2813521 & 12.77 & 0.27 &  6.92 & 2.42 \\
      & 2796 & 0.2814065 & 12.73 & 0.30 & 13.52 & 4.79 \\
      & 2803 & 0.2814064 & 12.73 & 0.30 & 13.52 & 4.79 \\
      & 2796 & 0.2817727 & 14.13 & 0.26 &  4.82 &  0.35\\
      & 2796 & 0.2818092 & 12.61 & 0.09 & 15.93 &  2.74\\
      & 2796 & 0.2819882 & 13.44 & 1.57 &  2.62 &  0.67\\
      & 2796 & 0.2820545 & 12.19 & 0.07 &  4.89 &  2.55\\
\end{supertabular}

\newpage
\tabcolsep1mm
\renewcommand{\arraystretch}{0.8}
\begin{center}
\tablecaption{\small \em Relative ratios for a sample of DLA 
at $z<1.5$. For each system, the observed ratios are shown 
in the first row, and 
the  ratios corrected for 
dust depletion 
are given in the second row.
\label{ratios}}
\tablehead{\hline}
\tabletail{\hline}
\begin{supertabular}{|c|c|ccccccccc|}
\small $z_{DLA}$ &\small  QSO &\small [Zn/Cr] &\small [Zn/Fe] &\small [Si/Fe] &\small [Ti/Fe] &\small [Mn/Fe] &\small [Ni/Fe]
&\small [Cr/Fe] &\small [S/Fe] &\small Ref.$^\ast$ \\ 
\hline\hline

\footnotesize 1.15 &\footnotesize \heb&\footnotesize $>1.04$
&\footnotesize 0.66 &\footnotesize 0.39 &\footnotesize 0.00
&\footnotesize -0.40 &\footnotesize $<$-0.53 &\footnotesize $<$-0.38
&\footnotesize &\footnotesize 1\\ 
&\footnotesize &\footnotesize $>0.05$ &\footnotesize 0.00
&\footnotesize -0.03 &\footnotesize -0.01 &\footnotesize -0.07
&\footnotesize $<$-0.03 &\footnotesize $<$-0.05 &\footnotesize
&\footnotesize \\\hline
\footnotesize 0.68&\footnotesize \hea &\footnotesize &\footnotesize
$<$0.02 &\footnotesize 0.72 &\footnotesize -0.32 &\footnotesize -0.70
&\footnotesize 
$<$0.31 &\footnotesize &\footnotesize &\footnotesize 1\\

  &\footnotesize &\footnotesize &\footnotesize $<$0.00 &\footnotesize 0.69 &\footnotesize -0.29 &\footnotesize -0.60 &\footnotesize $<$0.29 &\footnotesize &\footnotesize &\footnotesize \\ \hline

\footnotesize 0.3953        &\footnotesize PKS1229$-$0210&\footnotesize  $>0.35$ &\footnotesize $>0.85$ &\footnotesize $>0.56$ &\footnotesize &\footnotesize $>0.49$ &\footnotesize $>0.02$ &\footnotesize &\footnotesize &\footnotesize 2\\
 &\footnotesize &\footnotesize $>$-0.10 &\footnotesize $>0.24$ &\footnotesize $>0.11$ &\footnotesize &\footnotesize $>0.14$ &\footnotesize $>0.01$ &\footnotesize  &\footnotesize &\footnotesize \\\hline
\footnotesize 0.6560 &\footnotesize 3C336 &\footnotesize &\footnotesize $<1.04$ &\footnotesize &\footnotesize &\footnotesize $<$-0.29 &\footnotesize &\footnotesize &\footnotesize &\footnotesize 3\\
        &\footnotesize       &\footnotesize  &\footnotesize$<0.00$ &\footnotesize &\footnotesize &\footnotesize $<$-0.04 &\footnotesize &\footnotesize &\footnotesize &\footnotesize \\\hline
\footnotesize 0.6920  &\footnotesize 3C286&\footnotesize  0.50
&\footnotesize 0.71 &\footnotesize &\footnotesize &\footnotesize $<$-0.29 &\footnotesize $<0.46$ &\footnotesize $0.21$ &\footnotesize &\footnotesize 4,2\\
        &\footnotesize &\footnotesize -0.05 &\footnotesize 0.07 &\footnotesize &\footnotesize &\footnotesize $<$-0.06 &\footnotesize $<0.01$ &\footnotesize 0.04 &\footnotesize &\footnotesize \\\hline
\footnotesize 0.8598  &\footnotesize PKS0454+0356 &\footnotesize 0.07 &\footnotesize -0.07 &\footnotesize &\footnotesize &\footnotesize -0.36 &\footnotesize -0.57 &\footnotesize -0.14 &\footnotesize &\footnotesize 5\\
        &\footnotesize &\footnotesize 0.08 &\footnotesize -0.06 &\footnotesize &\footnotesize &\footnotesize -0.38 &\footnotesize -0.63 &\footnotesize -0.15 &\footnotesize &\footnotesize \\\hline
\footnotesize 1.0095 &\footnotesize EX~0302$-$233&\footnotesize  0.39 &\footnotesize $<0.89$ &\footnotesize &\footnotesize &\footnotesize  &\footnotesize $<0.50$ &\footnotesize $<$0.50 &\footnotesize &\footnotesize 6,2\\
        &\footnotesize &\footnotesize -0.10 &\footnotesize $<0.24$ &\footnotesize &\footnotesize &\footnotesize  &\footnotesize $<0.15$ &\footnotesize $<0.14$ &\footnotesize &\footnotesize \\\hline
\footnotesize 1.1491 &\footnotesize Q~1351+318&\footnotesize  0.57
&\footnotesize 0.64 &\footnotesize 0.45 &\footnotesize &\footnotesize
-0.16 &\footnotesize $<$-0.41  &\footnotesize 0.07 &\footnotesize
&\footnotesize 9 \\
        &\footnotesize &\footnotesize 0.00 &\footnotesize 0.01 &\footnotesize 0.00 &\footnotesize &\footnotesize -0.05 &\footnotesize $<$-0.03  &\footnotesize 0.00 &\footnotesize &\footnotesize \\\hline
\footnotesize 1.2232 &\footnotesize Q~1247+267&\footnotesize  0.20 &\footnotesize 0.37 &\footnotesize 0.28 &\footnotesize &\footnotesize -0.19 &\footnotesize$< -0.48$ &\footnotesize 0.17 &\footnotesize &\footnotesize 9\\
        &\footnotesize &\footnotesize -0.05 &\footnotesize 0.07 &\footnotesize 0.07 &\footnotesize &\footnotesize -0.09 &\footnotesize$< -0.13$ &\footnotesize 0.07 &\footnotesize &\footnotesize \\\hline

\footnotesize 1.3726 &\footnotesize Q~0935+417&\footnotesize  0.10 &\footnotesize 0.29 &\footnotesize &\footnotesize &\footnotesize -0.39 &\footnotesize &\footnotesize 0.19 &\footnotesize &\footnotesize 7,8\\
        &\footnotesize &\footnotesize -0.06 &\footnotesize 0.09 &\footnotesize &\footnotesize &\footnotesize -0.19 &\footnotesize &\footnotesize 0.11 &\footnotesize &\footnotesize \\\hline
\footnotesize 1.4200 &\footnotesize Q~1354+258&\footnotesize  0.20 &\footnotesize 0.42 &\footnotesize 0.30 &\footnotesize &\footnotesize -0.29 &\footnotesize -0.65 &\footnotesize 0.22 &\footnotesize &\footnotesize 9\\
        &\footnotesize &\footnotesize -0.06 &\footnotesize 0.10 &\footnotesize 0.07 &\footnotesize &\footnotesize -0.11 &\footnotesize -0.15 &\footnotesize 0.10 &\footnotesize &\footnotesize \\\hline

\end{supertabular}
\end{center}
\normalsize
\tabcolsep3mm
\renewcommand{\arraystretch}{1}
\noindent
\footnotesize
$^\ast$
(1) This work, 
(2) Boiss\'{e} et al.\,1998, and ref. therein, 
(3) Steidel et al. 1997,
(4) Meyer \& York 1992,
(5) Lu et al. 1996,
(6) Pettini \& Bowen 1997,
(7) Meyer, Lanzetta \& Wolfe 1995,
(8) Pettini et al. 1997a, and
(9) Pettini et al. 1999.
\normalsize


\newpage

\tablecaption{C\I measuremets in DLAs and in the Galaxy. 
\label{tbl-7}}

\tablehead{\hline}
\tabletail{\hline}
\begin{supertabular}{cccccc}

\scriptsize

Sightline  &
$z$ &
logN(H\I)&
logN(C\I)&
[Zn/H]&
References$^\ast$ \\ \hline

2000-330 & 3.19 & 19.69 & $<12.90$ &        & 1 \\
0741+4741& 3.02 & 20.03 & $<12.12$& $<-0.79 $ & 8 \\
0528-250 & 2.81 & 21.28 & $<13.48$ &$-0.76$ & 2 \\
0201+365 & 2.46 & 20.8 & $13.23$   &$>0.56$ & 3 \\
2243-6031& 2.33 & 20.39 & $<12.55$& $-1.06 $ & 8 \\
PHL\, 957 & 2.31 & 21.40 & $<12.65$&$$      & 4 \\
0013-004 & 1.97 & 20.70 & $13.65 $ &        & 5 \\
1331+170 & 1.78 & 21.18 & $13.28$  &$-1.27$ & 6 \\
0215+015 & 1.35 & 19.80 & $13.90$  &        & 7 \\
0515-4414 & 1.15 & 20.45 & $13.68$ &$ -0.99$& 11 \\
0454+039 & 0.86 & 20.69 & $13.64$  &$-0.8$  & 9 \\
3C286 & 0.69 & 21.25 & $13.57$     &$-1.1$  & 9 \\
1122-1649 &0.68 & 20.45 & $<13.45$ &$<-1.3$ & 11 \\
$\zeta $ Orp & 0 & 21.15 & $15.34$ &        & 10 \\
$\zeta $ Pup  & 0 & 20.0 & $12.80$ &        & 10 \\
\end{supertabular}
\vspace{.5cm}

\noindent
$^\ast$(1)Meyer \& York 1987a, (2)Meyer \& York 1987b and
ref. therein, (3) Prochaska \& Wolfe 1996, (4)Meyer et al.\, 1986,
Chaffee et al.\ 1986, (5) Ge et al.\, 1997, (6) Chaffee et al.\, 1988
and ref, therein, (7) Blades et al. 1985, (8) de la Varga 1999, (9) Boiss\'{e}
et al.\,1998 (10) Morton 1978, (11) This paper.



\newpage

\newpage 
\newpage 

\newpage 

\newpage 

\end{document}